\documentclass[11pt]{article}
\usepackage[dvips]{graphicx}
\usepackage{amsmath,amsfonts}
\usepackage{amssymb}
\usepackage{epsfig,slashed}
\usepackage{cite}
\usepackage{color,comment}
\usepackage[hmargin=2cm,vmargin=3.5cm]{geometry}

\newcommand{\be}{\begin{equation}}
\newcommand{\ee}{\end{equation}}
\newcommand{\bea}{\begin{eqnarray}}
\newcommand{\eea}{\end{eqnarray}}
\newcommand{\mbb}{\mathbb}
\newcommand{\ti}{\times}
\newcommand{\half}{\frac{1}{2}}
\newcommand{\mc}{\mathcal}

\newcommand{\beqa}{\begin{eqnarray}}
\newcommand{\eeqa}{\end{eqnarray}}

\newcommand{\im}{\mathrm{Im\;}}
\newcommand{\re}{\mathrm{Re\;}}

\def\bra{\langle}
\def\ket{\rangle}

\def\ap{\alpha^{\prime}}

\def\beq{\begin{equation}}
\def\eeq{\end{equation}}
\def\mr#1{\mathrm{#1}}

\newcommand{\vt}{\vartheta_1}

\def\ov{\overline}

\newcommand{\tab}[4]{\vartheta\left[ \begin{array}{c}#1 \\ #2 \end{array} \right] (#3,#4)}

\newcommand{\nn}{\nonumber}

\begin{document}

\title{}
\author{}
\date{}
\thispagestyle{empty}

\begin{flushright}
\vspace{-3cm}
{\small CPHT-RR054.0710 \\
\small DESY 10-115 \\
\small OUTP-10/16P  }
\end{flushright}
\vspace{1cm}

\begin{center}
{\bf\LARGE
One-loop Yukawa Couplings in Local Models}

\vspace{1.5cm}

{\bf Joseph P. Conlon\;}$^{1,2}$
{\bf,\hspace{.2cm} Mark Goodsell\;}$^{3}$
{\bf\hspace{.1cm} and\hspace{.2cm} Eran Palti\;}$^{4}$
\vspace{1cm}

{\it
$^1$ Rudolf Peierls Center for Theoretical Physics, 1 Keble Road, \\
Oxford, OX1 3NP, United Kingdom\\
$^2$ Balliol College, Oxford, OX1 3BJ, United Kingdom \\
$^3$ Deutsches Elektronen-Synchrotron DESY, Notkestrasse 85, D-22603 Hamburg, Germany.\\
$^4$ Centre de Physique Th´eorique, Ecole Polytechnique, CNRS, 91128 Palaiseau, France. \\
\vspace{3mm}
j.conlon1@physics.ox.ac.uk, mark.goodsell@desy.de, eran.palti@cpht.polytechnique.fr
}

\vspace{1cm}

{\bf Abstract}
\end{center}
\vspace{-.5cm}

We calculate the one-loop Yukawa couplings and threshold corrections for supersymmetric local models of branes at singularities in type IIB string theory.
We compute the corrections coming both from wavefunction and vertex renormalisation.
The former comes in the IR from conventional field theory running and in the UV from threshold corrections
that cause it to run from the winding scale associated to the full Calabi-Yau volume.
The vertex correction is naively absent as it appears to correspond to superpotential renormalisation.
However, we find that while the Wilsonian superpotential is not renormalised there is a physical vertex correction
in the 1PI action associated to light particle loops.

\clearpage

\tableofcontents


\section{Introduction}

Loop corrections to effective actions play an important general role in physics. Supersymmetric theories are celebrated for their special behaviour at loop level and for the protection  of holomorphic properties against renormalisation.
Specifically, the superpotential does not get perturbatively renormalised and the gauge kinetic function is
perturbatively renormalised only at one-loop. The study of one-loop gauge couplings and their threshold corrections
has been carried out extensively in both field and string theory. In this context there is an important distinction between the Wilsonian
gauge kinetic function, renormalised only at one-loop, and the physical coupling, which is corrected at all orders.
The study of one-loop Yukawa couplings in string theory has been less extensive, and the distinction between holomorphic and
physical couplings less clearly drawn. Here holomorphy properties have been taken to imply that
Yukawa couplings only receive one-loop corrections from wavefunction renormalisation.

In this paper we study one-loop Yukawa couplings for supersymmetric models of branes at singularities in type IIB string theory. These models are local in the sense that for much of the calculation the global compact completion of the internal manifold does not play a role \cite{Aldazabal:2000sa}. Previous studies of one-loop Yukawa couplings in heterotic and type IIA settings have been performed
in \cite{Wadia, Antoniadis:1992pm,0412206,Abel:2006yk}. There are two important motivations for this work. The first is that the one-loop behaviour of running couplings in local models is especially interesting
with respect to the volume of the full global manifold. For gauge threshold corrections, studied in \cite{09014350,09061920,Conlon:2009qa},
gauge couplings run to a scale set by the global volume of the manifold, the so called winding scale, rather than the local string scale. This has important consequences both practically, as for the case of local GUT models it implies the string scale is an order of magnitude below the GUT scale, and conceptually, in the sense of understanding the extent to which a local model can be decoupled from the bulk.
 It is natural to investigate the same behaviour for Yukawa couplings, where indeed we find precisely this property,
  with UV threshold corrections implying that the Yukawas are renormalised up to the winding scale rather than the
  naive cutoff, the local string scale.

The second motivation arose during the calculation and involves the distinction between physical and holomorphic couplings.
 In \cite{West,JackJonesWest,West:1991qt} it was argued that for supersymmetric field theories
 with massless particles, in apparent contradiction to the non-renormalisation theorem, it is possible for superpotential operators to receive finite loop corrections that are not associated to wavefunction renormalisation. In the case of Yukawa couplings this corresponds to vertex renormalisation. The vertex correction is generated
  in the IR by integrating over light loop
momenta. While the Wilsonian superpotential - corresponding to an action at a scale $E$ with light modes yet to be integrated over -
is not renormalised, the vertex correction is a physical feature of the 1PI action.
We find that local models of branes at singularities (and we expect also more general intersecting brane constructions) fall within this class of theories. We demonstrate explicit one-loop vertex renormalisation from the world-sheet perspective thereby giving, to our knowledge, the first realisation of this effect in string theory.

The paper is structured as follows. In section \ref{sec:cftbuildblock}
we review the basic CFT building blocks that are needed for the computation. This section establishes notation and
convention, as well as reviewing bosonic and fermionic correlators on the torus and annulus. We also introduce the model that we study throughout this paper: fractional D3 branes on the $\mbb{C}^3/\mbb{Z}_4$ orbifold.
Section \ref{sec:thresholds} contains the main calculation. We first compute the one-loop Yukawas via
a 3-point scattering amplitude computation. This demonstrates the key renormalisation properties of the Yukawas but also contains some ambiguities regarding the off-shell continuation. We subsequently resolve these by performing a 4-point calculation that reduces to the 3-point amplitude in a certain limit. We describe the structure of the string diagram and how it generates both wavefunction and vertex renormalisation.
In section \ref{sec:summary} we present a summary and discussion of our results. The
appendix contains a warm-up calculation of gauge threshold corrections as well as various properties of $\vartheta$-functions.

\section{CFT Building Blocks}
\label{sec:cftbuildblock}

The computation of the amplitudes requires the evaluation of various CFT correlators between world-sheet fields that are introduced through the vertex operators. In this section we collate the relevant correlators and also other miscellaneous
CFT results that are used. An incomplete list of useful references for these CFT correlators are \cite{Polchinski,Friedan:1985ge,Atick:1986rs,Atick:1986ns,Atick:1987gy,0412206,Abel:2005qn,Bianchi:2006nf,Benakli:2008ub, Olly}. All the amplitudes evaluated in this paper are cylinder (annulus) amplitudes and so all correlators are on this topology. We therefore begin with a brief description of this geometry before
describing the relevant correlators.

The cylinder has a single real modulus $t$ and is paramterised by a complex coordinate $z$. The circles at each end of the cylinder are positioned at $\re{(z)}=0,\half$ and are parameterised by $0\leq \im{z}\leq \frac{t}{2}$. The long cylinder limit is given by $t\rightarrow 0$ and corresponds to the open string UV and the closed string IR. The long strip limit is $t \rightarrow \infty$ and gives the open string IR and closed string UV. There is a single conformal Killing vector corresponding to translations parallel to the boundary.

The target space coordinates are the real worldsheet bosons $x^{M}\left(z,\bar{z}\right)$ where $M=0,...,9$. We further decompose $x^{M}=\left\{x^{\mu},x^m\right\}$ with $\mu=0,..,3$ denoting external directions and $m=4,..,9$ denoting internal directions. It will also be useful to pair the directions into complex pairs and we define
\be
X^{i} = x^{2i-2} + i x^{2i-1} \;,
\ee
where $i=1,...,5$. To save on clatter we usually drop the indices on the coordinates unless needed and denote
\be
X = x_1 + ix_2 \;.
\ee

There are two basic boundary conditions that can be imposed at each end of the cylinder
\be
\hbox{Neumann: } \partial_n X (z, \bar{z}) \equiv \half(\partial + \bar{\partial}) X(z, \bar{z}) = 0,
\ee
\be
\hbox{Dirichlet: } \partial_t X (z, \bar{z}) \equiv \half(\partial - \bar{\partial}) X(z, \bar{z}) = 0.
\ee
We have also defined the normal and tangential derivatives. In principle we can consider different boundary conditions at each end of the annulus but since we only study models involving D3 branes we restrict either to NN or DD boundary conditions. Henceforth  we denote the coordinate dependence $X(z)$ without implying holomorphic properties.

The cylinder can be obtained from the torus by quotienting under the identification $z \to 1 - \bar{z}$, with boundaries at $z = 1 - \bar{z}$. This is useful for relating bosonic ($X(z,\bar{z})$) correlators on the torus to those on the cylinder. The method of images
can then be used to obtain the cylinder correlators by starting with torus correlators and adding an image field at $1- \bar{z}$ for any field at $z$. The sign of the image correlator is positive for Neumann boundary conditions and negative for Dirichlet boundary conditions. The torus modular parameter $\tau$ is related to the cylinder modulus by $\tau=\frac{it}{2}$.

\subsection{Vertex operators}
\label{sec:vertexop}

The amplitudes are  calculated by inserting the vertex operators of the appropriate pictures into the partition function integral. In this section we briefly summarise the expressions for the vertex operators. We also note that as we always
calculate cylinder amplitudes,
the ghost charge should be zero and the sum of all the vertex operator `pictures' should vanish.

The bosonic vertex operator for a four-dimensional scalar $\phi$ is given in the $(-1)$ picture as
\be
{\cal V}_{\phi}^{-1} \left(z\right) = t^a e^{-\phi} \psi^i e^{ik \cdot x} \left(z\right)\;.
\ee
Here $z$ denotes the point on the worldsheet at which the vertex operator is inserted (which we integrate over). The scalar
Chan-Paton wavefunction is denoted $t^a$ and the field $\phi$ is the ghost from bosonising the $(\beta, \gamma)$ CFT.
The field $\psi^i$ can be (locally) bosonised in terms of free fields $H_i$ so that
\be
\label{psiboson}
\psi^i = e^{iH_i(z)}\;.
\ee
Here $i$ labels the complex direction.
Note that this bosonisation is only valid locally as the $\psi^i$ correlators depend on the spin structure. However these
amplitudes (which we give in section \ref{sec:fermcorr} below) are fixed uniquely in terms of this local bosonisation.
For economy of notation we typically suppress the CP index and wavefunction so that for a four-dimensional scalar
we have the (-1)-picture vertex operator
\be
{\cal V}_{\phi}^{-1}\left(z\right) = e^{-\phi} \psi^i e^{ik \cdot x} \left(z\right)\;.
\ee
The four-dimensional gauge field vertex operator is given by
\be
{\cal V}_{A}^{-1} \left(z\right) = A^a e^{-\phi} \epsilon_{\mu} \psi^{\mu} e^{ik \cdot x} \left(z\right)\;.
\ee
Here again $\psi^{\mu}$ can be bosonised with H-charge of $\pm 1$ and $\epsilon_{\mu}$ is the polarisation vector of the gauge boson
satisifying $\epsilon \cdot k = 0$.

The fermion vertex operator in the $(-\half)$ picture is given by
\be
{\cal V}_{\lambda}^{-\half} \left(z\right) = \lambda^a e^{-\frac{\phi}{2}} S_{10} e^{ik \cdot x} \left(z\right)\;.
\ee
Here $S_{10}$ is the ten-dimensional spin field which can be locally bosonised to
\be
\label{spinbos}
S_{10} = \prod_{i=1}^5 e^{iq^iH_i} \;,
\ee
where the H-charges $q_i$ are given by the spin $\pm \half$ of the complex direction components of the spinor.

To bring the amplitude into the appropriate zero ghost charge picture we can change pictures following the
prescription of \cite{Friedan:1985ge} using
\be
{\cal V}^{i+1}\left(z\right) = \underset{{w\rightarrow z}}{\mr{lim}} e^{\phi(w, \bar{w})} \mr{T_F} \left(w\right) {\cal V}^{i}\left(z\right) \;,
\ee
where we have the picture changing operator
\be
\mr{T_F}\left(w\right) = \frac12 \left( \psi_i \partial \overline{X}^i\left(w\right)  + \overline{\psi}_i\partial X^i\left(w\right)  \right) \;.
\ee
In practice the picture changing is evaluated using the operator product expansions (OPE)
\bea
e^{iaH\left(w\right)}e^{ibH\left(z\right)} &=& \left(w-z\right)^{ab}e^{i\left(a+b\right)H\left(z\right)} + ... \;, \label{opespin}\\
e^{ia\phi\left(w\right)}e^{ib\phi\left(z\right)} &=& \left(w-z\right)^{-ab}e^{i\left(a+b\right)\phi\left(z\right)} + ... \;, \label{opeghost} \\
\partial X\left(w\right)e^{ikX(z)} &=& -\frac{i\alpha'}{2}k^+\left(w-z\right)^{-1} e^{ikX(z)} + \partial X(z) e^{ikX(z)} + \ldots \;, \label{opeboson} \\
\partial \overline{X}\left(w\right)e^{ikX(z)} &=& -\frac{i\alpha'}{2}k^-\left(w-z\right)^{-1} e^{ikX(z)} + \partial \bar{X}(z) e^{ikX(z)} +
\ldots \;,
\eea
where the ellipses denote less divergent terms. Terms of $\mc{O}(z-w)^{-1}$ are dropped in the picture-changing,
although there are contributing `derivative' terms by combining an $\mc{O}(z-w)^{-1}$ term from the
$\partial X e^{ikX}$ correlator with an $\mc{O}(z-w)$ term from a higher-order term in the $e^{iaH(z)} e^{ibH(w)}$ OPE.
Recall that the $H_i$ are free fields and so
only OPEs with the same direction are non-vanishing. We have also introduced the
notation of complex momenta $k^{\pm}=k^1\pm ik^2$ for any complex direction
(other than the first for which $k^{\pm}=\pm k^1 + k^2$ to match the Minkowski signature), and defined
\be
\label{kXcomplex}
kX\left(z\right) \equiv \half \left(k^+ \cdot \overline{X}\left(z\right) + k^- \cdot X\left(z\right)\right)\;,
\ee
so that in complex notation we can write
\be
k\cdot x\left(z\right) = k_i X^i\left(z\right) \;.
\ee

\subsection{Bosonic Correlators}
\label{sec:boscorr}

We first evaluate the bosonic correlators, namely those involving the worldsheet bosons $X(z,\bar{z})$.
 Since the bosons are free worldsheet fields, for a correlator to be non-vanishing it must involve the same complex directions. Therefore such a correlator can be labeled by the associated direction: correlators involving $X^{1,2}$ are labeled external, while $X^{3,4,5}$ are internal.
However for computing Yukawa couplings we only need evaluate external correlators with Neumann boundary conditions, although for completeness
we also give expressions for Dirichlet correlators.

\subsubsection{Internal untwisted quantum correlators}

The quantum bosonic correlator on the cylinder can be derived from that on the covering torus (denoted by a subscript ${\cal T}$) which reads
\be
\langle X(z) \overline{X}(w) \rangle_{\mr{{\cal T},Qu}} = -\alpha' \log |\vartheta_1 (z-w)|^2 + \frac{2\pi\alpha'}{\im{\tau}} \left(\im (z-w)\right)^2\;.
\label{torusuntwicorr}
\ee
Here $\tau$ is the torus modular parameter.
For comparison with expressions in  \cite{Polchinski} note that here $X$ is a complexified coordinate.
As only correlators involving the same directions are non-vanishing
\be
\langle X(z) X(w) \rangle_{\mr{{\cal T},Qu}} = 0 \;,
\ee
as the two real directions give equal contributions of opposite sign.
From (\ref{torusuntwicorr}) one can obtain correlators on the cylinder (denoted by a subscript ${\cal A}$) through use
of the method of images.
\be
\langle X(z) \overline{X}(w) \rangle_{\mr{{\cal A}}} =  \half \left[ \langle X(z) \overline{X}(w) \rangle_{\mr{{\cal T}}} \, \pm \, \langle X(1-\bar{z}) \overline{X}(w) \rangle_{\mr{{\cal T}}} \pm \langle X(z) \overline{X}(1 - \bar{w})\rangle_{\mr{{\cal T}}} + \langle X(1-\bar{z}) \overline{X}(1-\bar{w}) \rangle_{\mr{{\cal T}}} \right],
\ee
where the plus sign applies for Neumann boundary conditions and the minus sign applies for Dirichlet boundary conditions. We can write the Neumann and Dirichlet correlator explicitly as
\bea
\label{neumannxx}
\bra X (z) \ov{X} (w) \ket_{\mr{{\cal A},Qu}}^{\mr{N}} &=&  -\alpha' \left(
\log \left|\vartheta_1 (z - w) \right|^2 + \log \left| {\vartheta_1 (\ov{z} + w)}\right|^2\right) + \frac{8\pi\alpha'}{t} \left(\im (z-w)\right)^2 \;, \\
\label{dirichletxx}
\bra X (z) \ov{X} (w) \ket_{\mr{{\cal A},Qu}}^{\mr{D}} &=&  -\alpha' \left(
\log \left| \vartheta_1 (z - w) \right|^2 - \log \left| {\vartheta_1 (\ov{z} + w)}\right|^2 \right)\;.
\eea
Here we have used the relation $\tau=\frac{it}{2}$ for the modular parameters of the cylinder and the covering torus.
The Dirichlet correlator has no zero mode since the string center of mass is fixed, whereas for Neumann boundary conditions the string can take any position.

Vertex operator computations with the bosonic fields can involve not only
the bare fields but also their derivatives.
For Neumann boundary conditions the vertex operators involve tangential derivatives $\partial_t X$ whereas for Dirichlet boundary conditions vertex operators involve normal derivatives $\partial_n X$. The relevant correlators are
\bea
\label{Neumanntt}
\bra \partial_t X(z) \partial_t \ov{X}(w) \ket_{\mr{{\cal A},Qu}}^{N} &=&
-\frac{\alpha'}{2} \left( \partial_z \partial_w \log \vartheta_1 (z-w)+ \mr{c.c.} \right) + \frac{4 \pi \alpha'}{t} \;, \\
\label{Dirictt}
\bra \partial_n X(z) \partial_n \ov{X}(w) \ket_{\mr{{\cal A},Qu}}^{D} &=&
-\frac{\alpha'}{2} \left( \partial_z \partial_w \log \vartheta_1 (z-w) + \mr{c.c.} \right) \;.
\eea

\subsubsection{Momentum exponential correlators and pole structures}
\label{sec:momenexpopole}

We also encounter correlators involving exponentials $e^{ikX}$. These are most easily calculated using real coordinates $x^{M}$ and momenta $k^M$. The relevant correlator
\be
\bra \prod_i e^{ik_i \cdot x\left(z,\bar{z}\right)} \ket \;,
\ee
is evaluated by contracting the scalars using the real forms\footnote{These are simply related to the complex versions by a factor of $\half$.} of the cylinder correlators (\ref{neumannxx}) and (\ref{dirichletxx}).
In general this is given by
\be
\prod_{i<j} e^{- k_i \cdot k_j \mc{G}(z_i - z_j)},
\ee
where $\mc{G}(z_i - z_j)$ is the bosonic correlator.
However for much of our calculation we only require the Neumann correlator in the limit $z_i\rightarrow z_j$,
 when we can drop the zero mode piece of (\ref{neumannxx}).  This is given by
\be
\label{expexp1}
\bra \prod_i e^{ik_i \cdot x\left(z_i,\bar{z}_i\right)} \ket^{N}_{{\cal A}} = \prod_{i<j} \left| \frac{\vartheta_1\left(z_{ij}\right)}{\vartheta_1'(0)} \right|^{\alpha' k_i k_j} \;.
\ee
We may also write (\ref{expexp1}) in complex co-ordinates and momenta as
\be
\bra \prod_i e^{ik_iX\left(z_j\right)} \ket^{N}_{{\cal A}} = \prod_{i<j} \left| \frac{\vartheta_1\left(z_{ij}\right)}{\vartheta_1'(0)} \right|^{\frac{\alpha'}{2}\left(k^+_i k^-_j + k^-_i k^+_j \right)} \;. \label{momexpcorrcom}
\ee
where we recall that the complex notation $k_iX\left(z\right)$ is defined in (\ref{kXcomplex}).

Another correlator that we require is
\be
\bra \partial X(w) \prod_i e^{ik_iX\left(z_j\right)} \ket^{N}_{{\cal A}} = -i\alpha' \prod_{i<j} k_j^+ \frac{\vartheta'_1\left(w-z_j\right)}{\vartheta_1\left(w-z_j\right)} \left| \frac{\vartheta_1\left(z_{ij}\right)}{\vartheta_1'(0)} \right|^{\frac{\alpha'}{2}\left(k^+_i k^-_j + k^-_i k^+_j \right)} \;, \label{momexpdercorr}
\ee
which can be deduced by acting on (\ref{expexp1}) with a derivative.

At this point we discuss a principle which greatly simplifies our calculations. The important point is that to probe
non-derivative terms in the action we do not need to know the full amplitude but rather only the
 zero momentum limit $k_i\rightarrow 0$. Given this it seems naively that bosonic correlators such as (\ref{momexpdercorr}) vanish. However it is also possible to generate a pole in the amplitude which when combined with the correlator (\ref{expexp1}) can generate inverse powers of momenta that cancel against the positive momentum powers leaving a result that is non-vanishing in the zero momentum limit. To see this consider the amplitude factor
\be
\label{polecancmom}
{\cal A} \supset \underset{{k_1\cdot k_2\rightarrow 0}}{\mr{lim}} \left[ \left(k_1 \cdot k_2\right) \int dz_1 \left|\frac{\vartheta_1\left(z_1-z_2\right)}{\vartheta_1'\left(0\right)}\right|^{k_1\cdot k_2} \left(\frac{\vartheta'\left(0\right)}{\vartheta_1\left(z_1-z_2\right)}\right)\right] = \frac{\left(k_1 \cdot k_2\right)}{\left(k_1 \cdot k_2\right)} = 1\;,
\ee
where we have used
\be
\frac{\vartheta_1\left(z\right)}{\vartheta_1'\left(0\right)} = z + {\cal O}\left(z^3\right) \;.
\ee
The pole at $z_1=z_2$ has cancelled the vanishing momentum prefactor. In practice this means that
evaluating certain amplitudes can simply amount to analysing their pole structure.

\subsection{Fermionic and Ghost Correlators}
\label{sec:fermcorr}

The amplitudes also involve correlators of spin fields, which after bosonisation as in (\ref{spinbos}) correspond
 to correlators of $H$ fields.
This includes the case of the $\psi$ correlators which are spin fields with $\pm 1$ H charge.
The correlators depend on the spin structure, denoted by indices $\left(\alpha\beta\right)=\left\{\left(00\right),\left(10\right),\left(01\right),\left(11\right)\right\}$, and read
\be
\label{hchargecorr}
\bra \prod_i e^{ia_iH\left(z_i\right)} \ket = K_{\alpha\beta} \left[\prod_{i<j} \left(\frac{\vartheta_1\left(z_{ij}\right)}{\vartheta'_1\left(0\right)}\right)^{a_ia_j}  \right]\vartheta_{\alpha\beta}\left(\sum_i a_i z_i + \theta_I\right) \;,
\ee
where $\theta_I$ is the orbifold twist in torus $I$. The constants $K_{\alpha\beta}$ are determined for each amplitude by the factorisation limit. This amounts to taking the limit $z_i \rightarrow z_j$ for all $i,j$ so that the amplitude factorises to the field theory amplitude times the string partition function. The spin structure is then matched to that of the partition function.
Note that using (\ref{opespin}) we deduce that only correlators where the total $H$-charge is zero are non-vanishing. This is known as $H$-charge conservation. These correlators were derived by Atick and Sen by considering their OPEs with the stress tensor, giving a set
of differential equations that can be solved to obtain the correlator. The details can be found in
\cite{Atick:1986ns, Atick:1986rs, 0412206}.

The ghost correlators can be found by the same method \cite{Atick:1986ns, Atick:1986rs}. The resulting correlators are very similar to the fermionic
correlators except with signs and powers reversed,
\be
\label{ghostchargecorr}
\bra \prod_i e^{ia_i\phi\left(z_i\right)} \ket = K_{\alpha\beta} \left[\prod_{i<j} \left(\frac{\vartheta_1\left(z_{ij}\right)}{\vartheta'_1\left(0\right)}\right)^{-a_ia_j}  \right]\vartheta_{\alpha\beta}^{-1}\left(-\sum_i a_i z_i \right) \;.
\ee
Again, the factors $K_{\alpha\beta}$ are determined by factorisation onto the partition function limit.

\subsection{Partition functions}
\label{sec:partfuncpre}

In the $2,3,4$ spin structures - those involving $\vartheta_{00}, \vartheta_{01},$ and  $\vartheta_{10}$ - the partition functions for the non-compact dimensions are given as follows
\begin{align}
\mathrm{Bosonic} :& \frac{1}{\eta^4 (it)} \frac{1}{(4\pi^2 \ap t)^2}, \nonumber \\
\mathrm{Fermionic} :& \bigg(\frac{\vartheta_\nu (0)}{\eta (it)} \bigg)^2, \nonumber \\
bc\ \mathrm{ghosts} :& \eta^2 (it), \nonumber \\
\beta \gamma \ \mathrm{ghosts} :& \frac{\eta(it)}{\vartheta_\nu (0)}, \nonumber \\
\mathrm{Total} :& \frac{\vartheta_\nu (0)}{\eta^3 (it)}  \frac{1}{(4\pi^2 \ap t)^2}.
\end{align}
For the 1 spin structure, which involves $\vartheta_{11}$, the above expressions must be changed, and they become
\begin{align}
\mathrm{Bosonic} :& \frac{1}{\eta^4 (it)} \frac{1}{(4\pi^2 \ap t)^2}, \nonumber \\
\mathrm{Fermionic} :& \bigg(\eta^4 (it) \bigg)^2, \nonumber \\
bc\ \mathrm{ghosts} :& \eta^2 (it), \nonumber \\
\beta \gamma \ \mathrm{ghosts} :& \frac{1}{\eta^2 (it)}, \nonumber \\
\mathrm{Total} :& \frac{1}{(4\pi^2 \ap t)^2}.
\end{align}
which assumes that the zero modes in the fermionic sector are saturated.
If this is not the case that the partition function vanishes due to integrating over the fermionic zero modes.
Note that we require no additional insertions for the $\beta \gamma$ ghosts; their zero modes must be explicitly excluded.
In practice however the effect of the fermionic and ghost partition functions are already incorporated into the
correlators (\ref{hchargecorr}) and (\ref{ghostchargecorr}).

The partition function for one compact torus $I$  with twist $\theta_I \ne 0$ is\footnote{For the partition function derivation see
\cite{09014350,09061920} for example}
\beq
Z_{I} = (-2\sin \pi \theta_I) \frac{\vartheta_\nu (\theta_I)}{\vartheta_1 (\theta_I)}
\eeq
while for an untwisted torus of area $T_2$ and complex structure $U= U_1 + i U_2$ it is
\beq
Z_{I} =   Z(t) \times \left\{ \begin{array}{cc} \frac{\vartheta_\nu (0)}{\eta^3 (it/2)} & \nu = 2,3,4 \\  1 & \nu = 1 \end{array} \right.\eeq
where
\beq
Z(t) \equiv  \sum_{n,m=-\infty}^{\infty} \exp [ -t \frac{\pi T_2}{\ap U_2} | n+ U m |^2 ] .
\label{CompactUntwistedPF}
\eeq
This assumes that both ends of the string are attached to the same brane stack, hence there is a zero mode as $t\rightarrow \infty$. If one end is on a stack displaced from the first by a (complex) displacement $z$, then we should modify $| n + Um|^2 \rightarrow |n + Um + \frac{z}{2\pi} \sqrt{\frac{U_2}{T_2}}|^2$ and there is no such zero mode.
In the following we shall define $Z(t)$ to be equal to $1$ when there is no $N=2$ sector in the amplitude.

\subsection{The Model}
\label{sec:themodel}

The model we use to compute the Yukawa couplings is a $\mbb{Z}_4$ toroidal orbifold with fractional D3 branes on the fixed point singularities. For our purposes the local model is sufficient and we could equally well work in a non-compact setting $\mbb{C}^3/\mbb{Z}_4$. The particular compact completion will not affect our general results. For a concrete global realisation which also cancels $N=2$ tadpoles we refer to \cite{09014350}. For simplicity we do not introduce orientifolds, which means that a global $N=4$ tadpole remains uncancelled. However
as the $\mc{N}=4$ sector does not contribute to running couplings our calculations are not sensitive to this tadpole. Therefore, although strictly the model is incomplete, it is sufficient for our purposes.

We begin by describing the local properties of the model near a $\mbb{C}^3/\mbb{Z}_4$ singularity. This model has been previously studied in \cite{09014350} and orientifolded versions of it have been analysed in \cite{09061920}. The advantages of this model are that despite being very simple it still has chiral matter with running gauge couplings.
\begin{figure}
\begin{center}
\epsfig{file=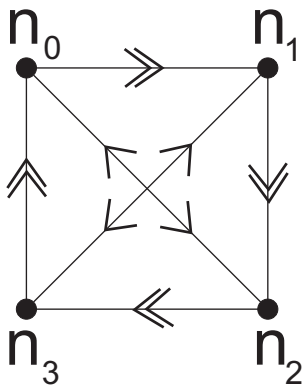,height=5cm}
\caption{The $\mathbb{Z}_4$ quiver. Each node $i$ has $n_i$ fractional branes on it, giving a $U(n_0) \ti U(n_1) \ti U(n_2) \ti U(n_3)$
gauge group. Each arrow corresponds to bifundamental chiral matter.}
\label{z4quiver}\end{center}\end{figure}
Locally the orbifold is $\mbb{C}^3/\mbb{Z}_4$, with the orbifold action $\Theta$ given by $\Theta:(z_1, z_2, z_3) \to (e^{2 \pi i/4} z_1,
e^{2 \pi i/4} z_2, e^{-2 \pi i/2} z_3)$. The orbifold twist vector is then $\frac14(a_1, a_2, a_3) = \frac{1}{4}(1,1,-2)$. The non-Abelian part of the
gauge group is $SU(n_0) \ti SU(n_1) \ti SU(n_2) \ti SU(n_3)$ and the spectrum is
\be
\sum_{i=0}^{3} \sum_{r=1}^3 \left(n_i,\bar{n}_{i+a_r} \right) \;,
\ee
where $\left(n_i,\bar{n}_{i+a_r} \right)$ denotes matter in the bifundemantal representation of $SU\left(n_i\right)\times SU\left(n_{i+a_r}\right)$.
The quiver diagram for the model is shown in figure \ref{z4quiver}. The superpotential is given by
\be
\label{superP}
W = \sum_{i=0}^3 \sum_{r,s,t=1}^3 \epsilon_{rst} \mathrm{Tr}\left( \Phi^r_{i,i+a_r}\Phi^s_{i+a_r,i+a_r+a_s}\Phi^t_{i+a_r+a_s,i} \right) \;,
\ee
where we define
\be
\Phi^r_{i,i+a_r}=\left(n_i,\bar{n}_{i+a_r}\right) \;.
\ee
The indices $r,s,t$ denote the plane that the bosonic field corresponds to (in terms of vertex operators this is equivalent
to the plane in which the boson has non-zero H charge). Local tadpole cancellation (equivalently cancellation of non-abelian anomalies) requires
\be
n_0=n_2\;,\;n_1=n_3\;,
\ee
and after imposing these the $\beta$ functions for the local gauge groups are given by
\be
\beta_{n_0} = \beta_{n_2} = -\beta_{n_1} = -\beta_{n_3} = \frac{1}{16\pi^2}(2n_1-2n_0) \;.
\ee
The Chan-Paton realisation of the orbifold twist is given for $N=1$ and $N=2$ sectors by
\bea
\Theta_{N=1} & = & \mathrm{diag\;}\left(1_{n_0},i_{n_1},-1_{n_2},-i_{n_3}\right)\;. \\
\Theta_{N=2} & = & \mathrm{diag\;}\left(1_{n_0},-1_{n_1},1_{n_2},-1_{n_3}\right)\;,
\eea
where $1_{n_i}$ corresponds to the unit $n_i\times n_i$ matrix. The embedding of the CP factors $\lambda_{n_in_j}$ of the gauginos and $\Phi^r$s into the full CP matrix of the singularity are given by
\bea
G^{1,2} &=& \mathrm{diag}\left(\lambda_{n_0n_0},\lambda_{n_1n_1},\lambda_{n_2n_2},\lambda_{n_3n_3}\right) \;, \nn \\
\Phi^{1,2} &=&
\left( \begin{array}{cccc}
0 & \lambda_{n_0n_1} & 0 &  0 \\
0 & 0 & \lambda_{n_1n_2} & 0 \\
0 & 0 & 0 & \lambda_{n_2n_3} \\
\lambda_{n_3n_0} & 0 & 0 &  0
 \end{array}\right) \;, \nn \\
\Phi^{3} &=&
\left( \begin{array}{cccc}
0 & 0 & \lambda_{n_0n_2}  &  0 \\
0 & 0 & 0 & \lambda_{n_1n_3} \\
\lambda_{n_2n_0} & 0 & 0 & 0 \\
0 & \lambda_{n_3n_1} & 0 &  0
 \end{array}\right) \;.
\eea
Note that the matrices satisfy the following
\be
\label{thetphicomm}
\Phi^i \Theta = \Theta \Phi^i e^{2\pi i \theta^i} \;.
\ee

\section{Yukawa threshold corrections}
\label{sec:thresholds}

In this section we address the main topic of the paper,
the calculation of one-loop Yukawa couplings. The approach we take is to calculate one-loop (annulus)
string scattering amplitudes which probe the Yukawa interaction in the theory. From the form of the amplitude we can deduce the resulting terms in the effective theory.

We begin with a discussion regarding general properties of the amplitudes and effective theory. Following this, in section \ref{sec:3ptamp} we calculate the 3-point amplitude that directly probes the Yukawa couplings at one-loop. The required physics is recovered from this calculation but there is an ambiguity regarding the on-shell limit. To resolve this ambiguity and check our results we perform a 4-point calculation in section \ref{sec:4ptamp} which reduces to the 3-point result in a particular limit.

\subsection{General structure}

There are two ways of computing the one-loop Yukawa coupling, via either a 3-point or a 4-point amplitude.
The Yukawa coupling does not involve derivatives and so we are only interested in terms independent of momentum.
The 3-point computation of $\langle \psi \psi \phi \rangle$ in principle should be performed at vanishing momentum. However in practice there are terms
of the form $\frac{k_i \cdot k_j}{k_i \cdot k_j}$ where numerator and denominator vanish on-shell. For this reason
it is necessary to compute with off-shell momenta and proceed on-shell only at the end of the computation.
This involves a need to resolve ambiguities associated with different ways to take the
 off-shell limit.

This ambiguity can be entirely resolved by going to a 4-fermion amplitude, and evaluating this in the limit where two of the fermion
vertex operators approach each other. This factorises the diagram onto a 3-point diagram with an off-shell scalar propagator.
The interaction of this with the two remaining fermions allows the
extraction of the Yukawa coupling. The 4-point amplitude can be computed with finite momenta and all particles on-shell, and so does not
involve any subtle questions of how to continue momenta off-shell. In practice both methods give the same structures, although the 4-point
computation is more rigorous as it is a purely on-shell computation.
\begin{figure}
\begin{center}
\epsfig{file=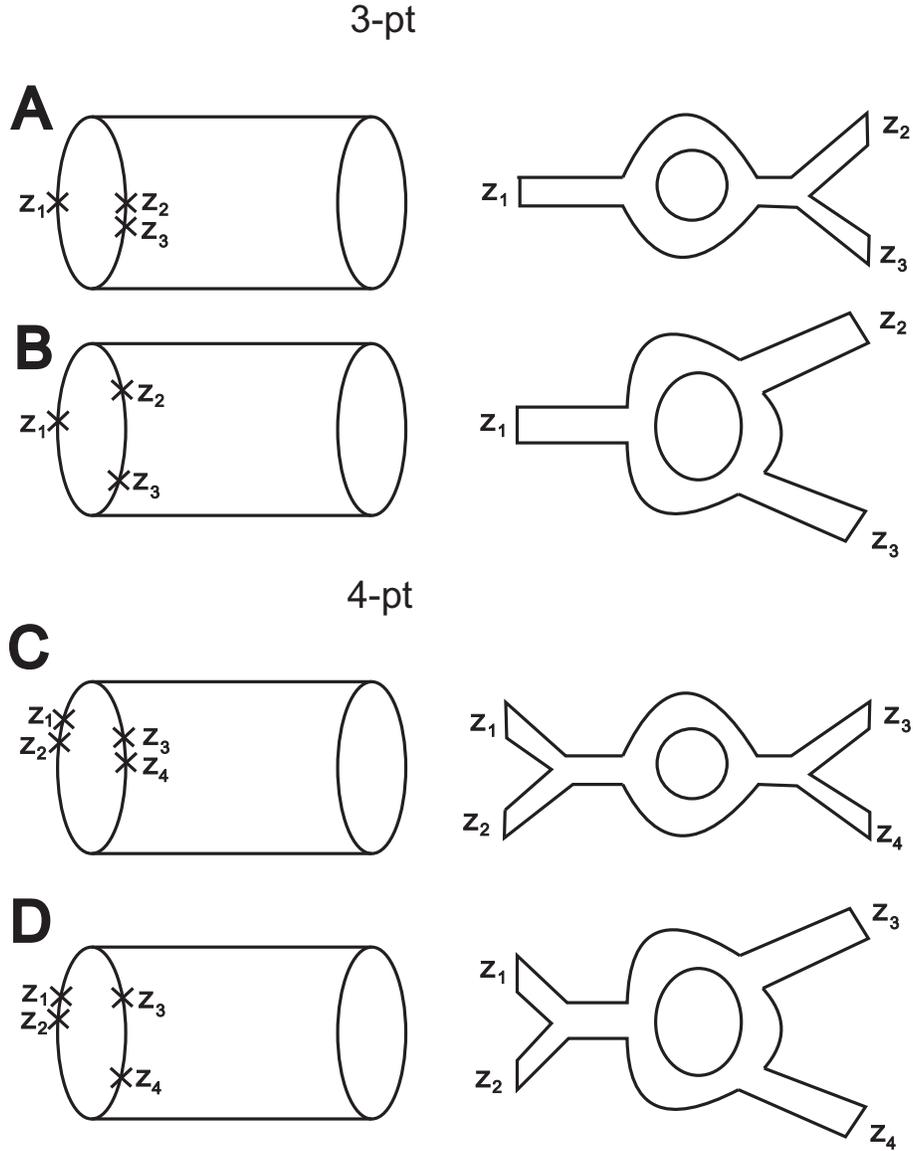, width=12cm}
\caption{The diagrams that enter the Yukawa coupling renormalisation for both 3 and 4-pt diagrams and their field theory limits.
Diagram A occurs in the limit that two of the vertex operators approach each other. This diagram then factorises onto a
wavefunction renormalisation diagram. Diagram B gives a limit where all vertex operators are well separated. In the field theory
this corresponds to a vertex correction to the Yukawa couplings. Diagrams C and D describe the same processes for the 4-point
function. In this case we always bring two of the fermionic vertex operators together in order to factorise onto the
scalar propagator. Again there are two types of correction to the Yukawa, one coming from wavefunction renormalisation and one
coming from a vertex correction.}
\label{BigDiagram}\end{center}\end{figure}

There are two basic set of contributions to the Yukawa couplings, labelled A and B (or C and D) in figure \ref{BigDiagram}.
The first (type A) comes from a loop-corrected propagator attached to a tree-level Yukawa vertex,
as illustrated by the field theory diagram.
In the context of a supersymmetric field theory this term is easily understood as coming from the one-loop correction to the K\"ahler
potential. Terms of type A receive contributions both from the infrared and the ultraviolet, associated to the running
of the kinetic term, and are logarithmically enhanced. We find that the appropriate running scale is given by $\ln (M_W^2/\mu^2)$. Here $\mu$ is the infrared cutoff - the energy scale of the process - whilst
$M_W$ is the winding scale of the compactification, given by $M_W = M_S R$ where $M_S$ is the string scale and $R$ is the bulk radius.
The infrared corresponds to standard field theory running with a cutoff at the probe energy scale.
In the ultraviolet the presence of the winding scale corresponds to stringy threshold effects. As for similar behaviour in gauge thresholds this is associated to a locally uncancelled tadpole in the closed string sector.

The term B is more subtle: from a field theory perspective, this corresponds to a pure vertex correction. Naively such a term should
be absent, as it appears to correspond to a renormalisation of the superpotential. However the non-renormalisation
theorems apply only to the Wilsonian superpotential, for which massless modes are still to be integrated over.
In fact we find term B is present. It arises in the infrared limit and should be understood as a renormalisation of the 1PI vertex -
the UV limit of the loop integral $(t \sim 0)$ gives no contribution.
For the 3-point function in the on-shell limit $k_i \cdot k_j \to 0$,
this term is generated in the strict $t \to \infty$ limit.
For the 4-point function with the scalar off-shell
by an amount $p^2 = s$, the correction is generated at $t \gtrsim 1/s$.
This term has an infrared Sudakov divergence for $t \gg 1/s$, which as for general vertex corrections
should be regulated by including an additional diagram with unobservable soft real emission.
The vertex correction therefore appears from integrating over loop momenta
at similar or smaller scales to the physics of the scattering process. This identifies this correction
as a genuinely infrared effect associated to the 1PI action, and for the case of massless particles
gives a finite one-loop correction to the zero-momentum Yukawa couplings. As this term comes from
integrating over light modes, it cannot be included in the Wilsonian superpotential defined at an energy scale $E$ (which is not renormalised).
However it does renormalise the `1PI superpotential' determining the physical couplings.

As an infrared effect, this should be present and calculable already in field theory.
This effect is indeed not unknown in the field theory literature \cite{JackJonesWest, West,West:1991qt},
although it does not appear to have wide circulation. For example, the massless Wess-Zumino model has a 2-loop renormalisation of the $\phi^3$ vertex \cite{JackJonesWest}. The effect arises through the presence of a term
\be
\int d^4 x \, d^4 \theta \, \frac{1}{\square} D^2 g(\Phi) + c.c \;,
\ee
where $g(\Phi)$ is a holomorphic function of the chiral superfield $\Phi$. Replacing $\int d^4 \theta$ by $\int d^2 \theta \bar{D}^2$ and using $\bar{D}^2 D^2 \Phi = \square \Phi$, this gives an effective contribution
\be
\int d^4 x \, d^2 \theta  \, g(\Phi) \;,
\ee
acting as an effective superpotential operator.  Such effects are important for all processes involving massless particles and in general for off-shell processes with energy scale $E$ and particle masses $m$ when $E^2/m^2 \gg 1$.

This effect is actually generic, and generated at one loop for any supersymmetric gauge theory with $N=1$ supersymmetry and no supersymmetric masses, i.e. for any theory described by the superpotential
\beq
W = \lambda_{ijk} \phi^i \phi^j \phi^k\;.
\eeq
Then there is a term generated in the 1PI action given by \cite{West:1991qt}
\begin{align}
\Gamma \supset& i \int\prod_{i=1}^3 \frac{d^4k_i}{(2\pi)^4}  \delta(\sum_i k_i) \int d^2 \theta \sum_s g_a^2 [T_i^{\phantom{i}d}(a) T_j^{\phantom{j}e}(a)]\lambda_{dek} \phi^i (-k_1) \phi^j (-k_2) \phi^k (-k_3) \nonumber \\
&\times \int_0^\infty dt \int_0^1 dx_1 \int_{0}^{x_2} dx_1 \, k_3^2 \exp [ -t (2 k_1 \cdot k_2 x_1 x_2 + k_1^2 x_1(1-x_1) + k_2^2 x_2(1-x_2) )]\;.
\label{RawWest}\end{align}

In the zero-momentum limit, taking $p_i \cdot p_j = -\frac{1}{2} u^2 + \frac{3}{2} \delta_{ij} u^2, u \rightarrow 0$ this gives a finite
Yukawa renormalisation
\begin{align}
\Gamma \supset& i \int d^2 \theta N \sum_a g_a^2 [T_i^{\phantom{i}d}(a) T_j^{\phantom{j}e}(a)]\lambda_{dek}  \phi^i \phi^j \phi^k \;,
\end{align}
where (in terms of Polygamma functions $\psi^\prime (z) \equiv \frac{d^2}{dz^2} \log \Gamma(z)$)
\begin{align}
N =&-\int_0^1 dx \frac{\log[x(1-x)]}{1-x(1-x)} \nn\\
=&\frac{1}{18}\bigg[ \psi^\prime (1/6) + \psi^\prime (1/3) - \psi^\prime (2/3)- \psi^\prime (5/6) \bigg]\nn\\
\approx& 2.344 \, .
\end{align}
Consider the model with quiver diagram \ref{z4quiver} and the Yukawa coupling between fields $(n_0,\bar{n}_1,0,0)$,$(0,n_1,\bar{n}_2,0)$,
$(\bar{n}_0,0,n_2,0)$. This then obtains a correction, for all $n_i > 1$, of
\begin{align}
\delta \lambda_{ijk} =&\frac{N}{2} (g_0^2 + g_1^2 + g_2^2) \lambda_{ijk} \nn\\
=& \frac{3N}{2 \mathrm{Im} S} \lambda_{ijk}.
\end{align}

We now compute the Yukawas, first the 3-point and then the 4-point amplitude.

\subsection{3-point amplitude}
\label{sec:3ptamp}

We start by recalling the structure of Yukawa couplings at tree level.
CFT computations of tree-level Yukawa couplings have been carried out for example in \cite{HamidiVafa, DFMS, 0302105, 0303083, Abel:2003vv, 0404134}.
We consider the $\mbb{C}^3/ \mbb{Z}_4$ orbifold and look at a Yukawa coupling associated to a triangle in the quiver of figure \ref{z4quiver} (for example ${\bf (n_0, \bar{n}_1)(n_1, \bar{n}_2)(n_2, \bar{n}_0)} $). Taking the first two fields to be fermions and the final field to
be bosonic, the `canonical' vertex operators are
\bea
\mc{V}_{-\half}^a(u_1, k_1, z_1) & = & t^a e^{-\phi/2} S^{\pm}(z_1) e^{i k_1 \cdot X (z_1)} e^{i q_1 \cdot H (z_1)}, \\
\mc{V}_{-\half}^b(u_2, k_2, z_2) & = & t^b e^{-\phi/2} S^{\mp}(z_2) e^{i k_2 \cdot X (z_2)} e^{i q_2 \cdot H (z_2)}, \\
\mc{V}_{-1}^c(u_3, k_3, z_3) & = & t^c e^{-\phi} e^{i k_3 \cdot X (z_3)} e^{i q_3 \cdot H (z_3)}.
\eea
The bosonised internal H-charges are
$q_1=(\half,-\half,-\half) \;,\; q_2=(-\half,\half,-\half) \;,\; q_3=(0,0,1) \;$,
and the overall H-charge structure of each operator is
\bea
\mc{V}_{-\half}^a & \sim &  |+ \, +>  \otimes | \, + \, - \, -> \nonumber \\
\mc{V}_{-\half}^a & \sim & |- \, ->  \otimes | \, - \, + \, -> \nonumber \\
\mc{V}_{-1}^c & \sim &  | \, 0 \, 0> \otimes | \, 0 \, 0 \, (++)> \;,
\eea
where we have split the external and internal H-charges.
We label the external spinors $e^{\pm \frac{i}{2} (H_1 + H_2)}$ as $S^+$ and $S^-$ appropriately; the above choice of H-charges
have sufficient generality to construct the full Lorentz-covariant amplitude.
The canonical vertex operators are appropriate for evaluating disk Yukawa couplings (ghost charge $-2$).
For the annulus we need to picture change two operators to give vanishing total ghost charge.
It is convenient to do this for one fermion and one boson, which we take to be $\mc{V}^b_{-\half}$ and $\mc{V}^c_{-1}$. The picture changed operators
are given by
\bea
\mc{V}_{\half}^b(w) & = & \lim_{z \to w} T(z) \mc{V}^b_{-\half}(w) =  \lim_{z \to w} e^{\phi} \partial X^{\mu} \psi_{\mu} (z) \mc{V}_{-\half}^b (w), \\
\mc{V}_{0}^c(w) & = & \lim_{z \to w} T(z) \mc{V}^c_{-1}(w) = \lim_{z \to w} e^{\phi} \partial X^{\mu} \psi_{\mu} (z) \mc{V}_{-1}^c (w).
\eea
Divergent terms of order $\mc{O}(z-w)^{-1}$ are dropped.\footnote{This comes from the origin of the picture changing in a contour integral,
$\int \frac{dz}{z-w} T(z) \mc{V}(w)$, for which only the pole term can contribute.}
The relevant annulus amplitude is
\bea
\mc{A} & = & \int \frac{dt}{t} \int dz_1 dz_2 dz_3 \mc{A}(z_1, z_2, z_3) \\
& = & \int \frac{dt}{t} \int dz_1 dz_2 dz_3 \left< V_{-1/2}^a\left(u_1,k_1,z_1\right)
\mc{V}_{\half}^b\left(u_2,k_2,z_2\right)
V_{0}^c\left(\varphi,k_3,z_3\right)
\right>. \nonumber
\eea
There are potentially three non-vanishing contributions to this amplitude.
The first contribution (denoted $\mc{A}_1$) comes from
picture changing in the internal directions, and the latter two ($\mc{A}_2$ and $\mc{A}_3$)
from picture changing in the external non-compact directions. Let us start by addressing the case of $\mc{A}_1$.

From the H-charges it is easy to see that internal picture changing can occur only
on the third torus, where the (-)(-)(++) structure
becomes (-)(+)(0) after picture changing. Due to the absence of internal momenta, there are also no derivative terms (i.e. terms where the derivative part of the PCO is contracted with the exponential).
Writing $q_2^{'} = (-\half, \half, \half)$, the resulting amplitude has the structure
\bea
\mc{A}_1(z_i) &=& \frac14 u_1 u_2 \varphi \mathrm{Tr}[CP] \left<e^{-\half\phi}(z_1)e^{\half\phi}(z_2)\right>
 \left< e^{\pm i\frac12(H_1+H_2)(z_1)} e^{\mp i\frac12(H_1+H_2)(z_2)} \right> \nonumber \\
& &  \left<e^{iq_1 \cdot H}(z_1) e^{i q^{'}_2 \cdot H}(z_2) \right>
 \left< e^{i k_1 \cdot X (z_1)} e^{i k_2 \cdot X (z_2)} e^{i k_3 \cdot X (z_3)} \right>
 \left< \partial \bar{X}^3 (z_2) \partial X^3 (z_3) \right>.
\eea
Here $u_1$ and $u_2$ denote the fermions and $\varphi$ the boson, and $\mathrm{Tr}[CP]$ denotes a trace over the CP factors.
As the bosonic correlators are independent of spin structure the only
spin structure dependence comes from the ghost and fermion
correlators. These are all of the form
$$
\left<e^{-\half\phi}(z_1)e^{\half\phi}(z_2)\right>, \quad \left<e^{\pm \half H_i}(z_1)e^{\mp \half H_i}(z_2)\right>,
$$
where $i$ runs over both internal and external coordinates, and are easy to evaluate using the results of section \ref{sec:fermcorr}.
The combined result for fermion plus ghost correlators in the $\alpha \beta$ spin structure is
\be
\mc{A}_{1,\alpha \beta} = \eta_{\alpha \beta} \left( \frac{\vartheta_1(z_1 - z_2)}{\vartheta^{'}(0)} \right)^{-1} \vartheta_{\alpha \beta}\left(\frac{z_1-z_2}{2}\right)
\prod_{i=1}^3 \vartheta_{\alpha \beta}  (q_1^{i} z + q'^{i}_2 w + \theta_i)\;.
\ee
The sum over spin structures is simplified by the Riemann identity
\bea
& &\sum_{\alpha\beta} \eta_{\alpha\beta} \vartheta_{\alpha\beta}(x)\vartheta_{\alpha\beta}(y)\vartheta_{\alpha\beta}(u)\vartheta_{\alpha\beta}(v) = 2\vartheta_1(x')\vartheta_1(y')\vartheta_1(u')\vartheta_1(v') \;, \nn \\
\hbox{ where }& &x' = \frac12(x+y+u+v) \;,\; y' = \frac12(x-y+u-v) \;,\;\nn \\
& &u' = \frac12(x+y-u-v) \;,\;v' = \frac12(x-y-u+v) \;.
\eea
This gives
\be
\mc{A}_1 \sim \vartheta_1(\theta_1 + \theta_2 + \theta_3)\vartheta_1(\theta_2) \vartheta(z_1 - z_2 + \theta_1) \vartheta_1(\theta_3) \;,
\ee
which vanishes as $\theta_1 + \theta_2 + \theta_3 = 0$ for supersymmetric orbifolds.
As this term potentially gets contributions from all values of the annular
modular parameter $t$ we associate it with renormalisation of the Wilsonian superpotential which, as expected, vanishes in a supersymmetric theory.

We next consider contributions to the amplitude that come from picture changing operators acting on the external coordinates.
In the external directions the canonical operators have H-charges $(+, +)(-, -)(0, 0)$.
Up to overall flips in the H-charges and the choice of direction on which to act, there are essentially two distinct
ways of picture changing these operators: first to $(+, +)(-, ---), (0, ++)$ and secondly to $(+, +)(-, +)(0, --)$. There are also the analogous picture changing choices for the first complex extremal direction which serve to complete the momentum structure into a Lorentz covariant one. We denote the first amplitude by $\mc{A}_2$ and the second amplitude by $\mc{A}_3$.

The $\mc{A}_2$ amplitude has the following structure
\bea
\mc{A}_3 &=& \frac14 u_1 u_2 k_2^{2+} k_3^{2-} \varphi \mathrm{Tr}[CP] \left<e^{-\half\phi}(z_1)e^{\half\phi}(z_2)\right>
 \left< e^{i \frac{H_1}{2} (z_1)} e^{-i \frac{3H_1}{2}(z_2)} e^{i H_1(z_3)} \right>
 \nonumber \\
& &   \left< e^{i \frac{H_2}{2} (z_2)} e^{-i \frac{H_2}{2}(z_2)} \right>
\left<e^{iq_1 \cdot H (z_1)} e^{i q_2 \cdot H (z_2)} e^{i q_3 \cdot H (z_3)} \right>
 \left< e^{i k_1 \cdot X (z_1)} e^{i k_2 \cdot X (z_2)} e^{i k_3 \cdot X (z_3)} \right>.
\eea
The factors of $k_2^{2+}$ and $k_3^{2-}$ arise from the picture changing contraction with $e^{ik \cdot X}$. From picture changing applied
in the first complex plane there is also a similar term $k_2^{1+} k_3^{1-}$, but note that even with this contribution the $\mc{A}_2$ amplitude by itself cannot form a Lorentz covariant quantity.
As the amplitude has an overall momentum prefactor of $k_2^{2+} k_3^{2-}$, it naively appears to
 vanish at zero momentum. However, this prefactor turns out to be cancelled by an integral over the
vertex operator insertions, which give a momentum pole that cancels this.
Using (\ref{hchargecorr}) we get
\be
\left< e^{i \frac{H}{2}(z_1)} e^{-i \frac{3H_1}{2}(z_2)} e^{i H_1(z_3)} \right>  = \left(
\frac{\vartheta_1(z_1 - z_2)}{\vartheta_1^{'}(0)} \right)^{-3/4} \left( \frac{\vartheta_1(z_1 - z_3)}{\vartheta_1^{'}(0)} \right)^{\half}
\left( \frac{\vartheta_1(z_2 - z_3)}{\vartheta_1^{'}(0)} \right)^{-3/2}
\vartheta_{\alpha \beta}  \left(\frac{z_1 - 3z_2}{2} + z_3 \right). \nonumber
\ee
Combining this with the other correlators we obtain for the combined fermion and ghost correlator
\bea
\mc{A}_2 & \sim & \sum_{\alpha \beta} \delta_{\alpha \beta} \, \eta^3 \left( \frac{\vartheta_1(z_1 - z_2)}{\vartheta_1^{'}(0)} \right)^{-1} \left( \frac{\vartheta_1(z_2 - z_3)}{\vartheta_1^{'}(0)} \right)^{-2} \ti \\
& & \vartheta_{\alpha \beta} \left(\frac{z_1 - 3z_2}{2}+ z_3 \right)
\vartheta_{\alpha \beta} \left(\frac{z_1 - z_2}{2} + \theta_1 \right)
\vartheta_{\alpha \beta} \left(\frac{-z_1 + z_2}{2} + \theta_2 \right)
\vartheta_{\alpha \beta}\left(\frac{- z_1 - z_2}{2} + z_3 + \theta_3 \right).
\nonumber
\eea
Summing over spin structures using the Riemann identity gives
\be
\label{eq40}
\mc{A}_3 \sim \eta^3  \left( \frac{\vartheta_1(z_1 - z_2)}{\vartheta_1^{'}(0)} \right)^{-1} \left( \frac{\vartheta_1(z_2 - z_3)}{\vartheta_1^{'}(0)} \right)^{-2} \vartheta_1  \left( -z_2 + z_3 \right)
\vartheta_1 \left( z_1 - z_2 + \theta_1  \right)
\vartheta_1 \left( \theta_2 \right)
\vartheta_1 \left( z_2 - z_2 + \theta_3 \right).
\ee
Note that eq. (\ref{eq40}) incorporates the fermionic and ghost partition functions (as it factorises correctly as
$z_1 \to z_2 \to z_3$) but does not include the bosonic partition functions. The spare $\eta^3$ comes from the ghost correlator and will
cancel a similar term from the bosonic partition function.

The bosonic correlator gives
\be
\langle e^{i k_1 \cdot X (z_1)} e^{i k_2 \cdot X (z_2)} e^{i k_3 \cdot X (z_3)} \rangle = \nonumber
\ee
\be
 \frac{\prod_i \left( - 2 \sin \pi \theta_i \right) \exp \left[ -k_1 \cdot k_2 \mc{G}(z_1 - z_2) - k_1 \cdot k_3 \mc{G}(z_1 - z_3) - k_2 \cdot k_3 \mc{G}(z_2 - z_3) \right] }
{\vartheta_1 (\theta_1 ) \ti \vartheta_1 (\theta_2 ) \ti \vartheta_1 (\theta_3 )} Z(t)
\ee
The denonimator comes from the bosonic partition function, ensuring that the amplitude behaves correctly in the
$z_1 \to z_2 \to z_3$ limit. Note that as all momenta are aligned along spacetime directions the $X_i$ are purely in the
external directions: in the internal directions we obtain simply the bosonic partition functions.

We temporarily replace the full correlator $\mc{G}(z_i - z_j)$ by its $\vartheta_1$ part (we will justify this retrospectively later)
to get for the combined - fermion, ghost and boson - correlator
\bea
\label{eq41}
\mc{A}_2(z_i) & = & \prod_i (-2 \sin \pi \theta_i)  \\
& \ti &
(k_2^{2+} k_3^{2-}) \left( \frac{\vartheta_1(z_1 - z_2)}{\vartheta_1^{'}(0)} \right)^{-1 + k_1 \cdot k_2}
\left( \frac{\vartheta_1(z_1 - z_3)}{\vartheta_1^{'}(0)} \right)^{k_1 \cdot k_3}
\left( \frac{\vartheta_1(z_2 - z_3)}{\vartheta_1^{'}(0)} \right)^{-2 + k_2 \cdot k_3}
\nonumber \\
& \ti &
\left( \frac{\vartheta_1(-z_2 + z_3)}{\vartheta_1^{'}(0)} \right)
\left( \frac{\vartheta_1(z_1 - z_2 + \theta_1)}{\vartheta_1(\theta_1)} \right)
\left( \frac{\vartheta_1(\theta_2)}{\vartheta_1(\theta_2)} \right)
\left( \frac{\vartheta_1(z_3 - z_2 + \theta_3)}{\vartheta_1(\theta_3)} \right) Z(t).
\nonumber
\eea
Note this has a pole as $z_1 \to z_2$ and integrating over vertex operator loci will give a pole in $k_1 \cdot k_2$ as discussed in section \ref{sec:momenexpopole} above.
We also note that if $\theta_i = 0$ (as occurs in $\mc{N}=2$ or $\mc{N}=4$ sectors) then the $\vartheta_1(\theta)$ terms in the denominator (not numerator) should be replaced by $\eta^3$. To obtain the full amplitude, we need
to integrate (\ref{eq41}) over the annular modular parameter and the vertex operator loci and include the
Chan-Paton factors:
\be
\mc{A}_2 = \int \frac{dt}{t (8 \pi^2 \alpha' t)^2} \int dz_1 dz_2 dz_3 \mc{A}_2(z_1, z_2, z_3) \hbox{Tr}_L(t^1 t^2 t^3 \theta^K) \hbox{Tr}_R (\theta^K).
\label{babel}
\ee
(We include the factor of $t^{-2}$ coming from integration over the non-compact momenta).

We now study the form
of eq (\ref{eq41}) for different sectors.
For $\mc{N}=4$ sectors eq. (\ref{eq41}) vanishes trivially as there is a $\vartheta_1(\theta_2)$ in the numerator which vanishes for
$\theta_2 = 0$. For $\mc{N}=1$ there are two poles at $z_2 = z_3$ and also at $z_1 = z_2$.
So when integrating over the vertex operator coordinates we will pick up a double pole overall,
and the amplitude will look like
$$
\int \frac{dt}{t^2} Z(t) \hbox{Tr}_L(t^1 t^2 t^3 \theta^K) \hbox{Tr}_R (\theta^K) \frac{k_2^{2+} \cdot k_3^{2-}}{(k_1 \cdot k_2)(k_2 \cdot k_3)}.
$$
This diverges in the on-shell limit $k \to 0$ and it is not clear how to interpret this. Fortunately it is not necessary to do so
as the Chan-Paton traces vanish for the $\mc{N}=1$ sector when we trace over the right hand side of the string. This is similar to the
 $\mc{N}=1$ sector for gauge threshold corrections where a divergent term is cancelled by a vanishing Chan-Paton trace.

Now consider the $\mc{N}=2$ sector for which $\theta_3$ is an integer. In this case $\vartheta_1(z_2 - z_2 + \theta_3) \to 0$
as $z_3 \to z_2$ as an integer shift does not affect the zeros of $\vartheta_1$. So we now have \emph{two} positive powers
of $(z_2 - z_3)$ as $z_2 \to z_3$, thereby cancelling the pole in $(z_2 - z_3)$. We can then write the amplitude as
\bea
\label{dfg}
\mc{A}_2 & = & \int \frac{dt}{t^3} Z(t) \int dz_1 dz_2 dz_3 \hbox{Tr}_L(t^1 t^2 t^3 \theta^K) \hbox{Tr}_R (\theta^K) \prod_i \left( - 2 \sin \pi \theta_i \right) \\
& & (k_2^{2+} \cdot k_3^{2-}) \left( \frac{\vartheta_1(z_1 - z_2)}{\vartheta_1^{'}(0)} \right)^{-1 + k_1 \cdot k_2}
\left( \frac{\vartheta_1(z_1 - z_3)}{\vartheta_1^{'}(0)} \right)^{k_1 \cdot k_3}
\left( \frac{\vartheta_1(z_2 - z_3)}{\vartheta_1^{'}(0)} \right)^{k_2 \cdot k_3} \ti \nonumber \\
& & \left( \frac{\vartheta_1(z_1 - z_2 + \theta_1)}{\vartheta_1(\theta_1)} \right)
\left( \frac{\vartheta_1(\theta_2)}{\vartheta_1(\theta_2)} \right)
\left( \frac{\vartheta_1^{'}(0)}{\eta^3} \right). \nonumber
\eea
where we have used the fact that we are in an $\mc{N}=2$ sector and so $\theta_3 \in \mbb{Z}$. Before we evaluate this
let us consider the $\mc{A}_3$ amplitude.

The $\mc{A}_3$ amplitude involves the picture changing of a fermionic field from $-1/2$ to $+1/2$ H-charge.
This means we need to study the derivative terms and in particular evaluate the
amplitude before taking the picture-changing limit. The H-charges in this limit are
\bea
\psi^{-1/2}_1(z_1) &=& \half(+,+,+,-,-) \;, \nn \\
\psi^{-1/2}_2(z_2) &=& \half(-,-,-,+,-) \;, \nn \\
PCO^{+1}(w) &=& (++,0,0,0,0) \;, \nn \\
\phi^{0}(u) &=& (--,0,0,0,++) \;.
\eea
Here $\psi$ denote the fermions located at $z_1$ and $z_2$, $\phi$ the boson located at $u$ and $PCO^{+1}(w)$ is the location of the picture changing field, and we are interested in the limit $w \to z_2$.
Using (\ref{hchargecorr}) we can evaluate the spin and ghost correlators. For spin structure dependent terms we get
\be
\frac{\vartheta_{\alpha \beta} \left( \frac{z_1-z_2}{2} + w - u \right) \vartheta_{\alpha \beta} \left( \frac{z_1 - z_2}{2} \right)
\vartheta_{\alpha \beta} \left( \frac{z_1-z_2}{2} + \theta_1 \right) \vartheta_{\alpha \beta} \left( \frac{-z_1 + z_2}{2} + \theta_2 \right)
 \vartheta_{\alpha \beta} \left( \frac{-z_1-z_2}{2} + u + \theta_3 \right)}{\vartheta_{\alpha \beta}(\frac{z_1 + z_2}{2} - w)}
 \ee
\be
= \frac{\vartheta_{\alpha \beta} \left( \frac{-z_1+z_2}{2} - w + u \right) \vartheta_{\alpha \beta} \left( \frac{z_1 - z_2}{2} \right)
\vartheta_{\alpha \beta} \left( \frac{-z_1+z_2}{2} - \theta_1 \right) \vartheta_{\alpha \beta} \left( \frac{z_1 - z_2}{2} - \theta_2 \right)
 \vartheta_{\alpha \beta} \left( \frac{z_1+z_2}{2} - u - \theta_3 \right)}{\vartheta_{\alpha \beta}(\frac{-z_1 - z_2}{2} + w)}.
\ee
We can now use the 5-theta identity (\ref{FiveTheta}) to simplify this to
\be
\label{xgun}
\frac{\vartheta_1 \left( -w + u + \theta_3 \right) \vartheta_1 \left( z_2 - w + \theta_2 \right)
\vartheta_1 \left( z_1 - w + \theta_1 \right) \vartheta_1 \left( z_2 - w \right)
 \vartheta_1 \left( z_1 - u \right)}{\vartheta_1 (z_1 + z_2 - 2w)}.
\ee
The spin-structure independent part is
\be
\label{ygun}
\vartheta_1(z_1 - z_2)^{-1} \vartheta_1(z_1 - w) \vartheta_1(z_1 - u)^{-1} \vartheta_1(w - u)^{-1}.
\ee
Combining (\ref{xgun}) and (\ref{ygun})
we have a zero as $w \to z_2$, and so to cancel this we need to take the pole from
$$
\partial X^{1-} (w) e^{ik_2 \cdot X(z_2)} \to \frac{i k_2^{1-}}{(z_2 - w)} e^{i k_2 \cdot X(z_2)}.
$$
We then obtain overall in the limit that $(w - z_2) \to 0$
\be
k_2^{1-} k_3^{1+} \frac{2 \vartheta_1(u - z_2 + \theta_3) \vartheta_1(\theta_2) \vartheta_1(z_1 - z_2 + \theta_1) \vartheta_1(z_1 - u)}
{\vartheta_1(z_1 - z_2) \vartheta_1(z_1 - u) \vartheta_1(z_2 - u)}.
\ee
The amplitudes vanishes for both $\mc{N}=4$ and $\mc{N}=1$ sectors. In the $\mc{N}=2$ case, we can take $\theta_3 = 0$, obtaining
\be
\int \frac{dt}{t^3} \int dz_1 dz_2 dz_3 k_2^{1-} k_3^{1+} \hbox{Tr}_L(t^1 t^2 t^3 \theta^K) \hbox{Tr}_R (\theta^K) \prod_i \left( - 2 \sin \pi \theta_i \right)
\frac{\vartheta_1(z_1-z_2 + \theta_1) \eta^3}{\vartheta_1(z_1-z_2) \vartheta_1(\theta_1)} Z(t).
\ee
We can combine with the $\mc{A}_2$ case of (\ref{dfg}) including the equal contributions from both the complex external directions to give
\be
\int \frac{dt}{t^3} \int dz_1 dz_2 dz_3 \, \left(k_2 \cdot k_3\right) \hbox{Tr}_L(t^1 t^2 t^3 \theta^K) \hbox{Tr}_R (\theta^K) \prod_i \left( - 2 \sin \pi \theta_i \right)
\frac{\vartheta_1(z_1-z_2 + \theta_1) \eta^3}{\vartheta_1(z_1-z_2) \vartheta_1(\theta_1)} \langle \prod e^{ik \cdot X} \rangle.
\ee
This now gives a Lorentz-covariant structure.

There are two interesting limits in this amplitude. Overall we want a contribution that is independent of momentum and so it is necessary
to cancel the $k_2 \cdot k_3$ prefactor in the amplitude, which we can do in two ways. First, we can take $z_1 \to z_2$, where there
is a pole in $(z_1 - z_2)$ that is regulated by the bosonic correlator, giving a pole in $k_1 \cdot k_2$. Secondly, we can consider the limit
where $t \gg 1$, $|z_i - z_j| \gg 1$, for which the integrand becomes a constant. The integral then looks like $\sim
k_2 \cdot k_3 \int \frac{dt}{t^3} dz_1 dz_2 dz_3 \langle e^{ik \cdot X} \rangle$. This is regulated by the bosonic correlators in the
limit that $t \sim \frac{1}{k_i \cdot k_j}$.

We first consider the case of $z_1 \to z_2$, when the amplitude looks like
\be
\int d(z_1 - z_2) \left( \frac{\vartheta_1(z_1 - z_2)}{\vartheta_1^{'}(0)} \right)^{-1 + k_1 \cdot k_2}
\sim \int d(z_1 - z_2) (z_1 - z_2)^{-1 + k_1 \cdot k_2} \to_{k_1 \cdot k_2 \to 0} \frac{1}{k_1 \cdot k_2}.
\ee
We perform this integral, pick up the pole and put $z_1 = z_2$ in the rest of the amplitude, leaving
\bea
\mc{A}_2 & = & \int \frac{dt}{t^3} \int dz_2 dz_3 \hbox{Tr}_L(t^1 t^2 t^3 \theta^K) \hbox{Tr}_R (\theta^K) \prod_i \left( - 2 \sin \pi \theta_i \right) \nonumber \\
& & \frac{(k_2 \cdot k_3)}{(k_1 \cdot k_2)} \left( \frac{\vartheta_1(z_2 - z_3)}{\vartheta_1^{'}(0)} \right)^{(k_1 + k_2) \cdot k_3}
\left( \frac{\vartheta_1( \theta_1)}{\vartheta_1(\theta_1)} \right)
\left( \frac{\vartheta_1(\theta_2)}{\vartheta_1(\theta_2)} \right)
\left( \frac{\eta^3}{\eta^3} \right) Z(t), \nonumber
\eea
where we have used $\vartheta_1^{'}(0) = \eta^3$. For the limit $z_1 \to z_2$ the effective restriction of the integral to $z_1 = z_2$ in order to pick up the pole allows the use of the $\vartheta_1$ approximation to $\mc{G}(z_{ij})$ - we only need care about the behaviour of $\mc{G}(z_{ij})$
in the vicinity of $z_1 = z_2$, which is captured by the $\vartheta_1$ approximation. In the on-shell limit of $k_i \to 0, \sum k_i = 0, k_i^2 = 0$ the remaining $\vartheta$ functions and
momentum factors all either drop out or cancel, giving
$$
\int \frac{dt}{t^3} \int dz_2 dz_3 \frac{k_2 \cdot k_3}{k_1 \cdot k_2} \hbox{Tr}_L(t^1 t^2 t^3 \theta^K) \hbox{Tr}_R (\theta^K)  \prod_i \left( - 2 \sin \pi \theta_i \right) Z(t).
$$
The $z$ integrals are now trivial and give a factor of $t^2$, giving overall
\be
\label{wywy}
\mc{A} \sim \int \frac{dt}{t} \frac{k_2 \cdot k_3}{k_1 \cdot k_2}
\hbox{Tr}_L(t^1 t^2 t^3 \theta^K) \hbox{Tr}_R (\theta^K)  \prod_i \left( - 2 \sin \pi \theta_i \right) Z(t),
\ee
which is very similar to the structure that emerged from the study of gauge threshold corrections presented in the appendix. However one slightly troubling aspect of this expression is that numerical evaluation of it depends on the off-shell prescription: it is clear $\frac{k_2 \cdot k_3}{k_1 \cdot k_2}$
should cancel, but the precise numerical coefficient is not unambiguously determined. This ambiguity can be removed by performing a
4-point computation.

Nonetheless, the essence of this expression is that it is
simply proportional to the propagator for the $\Phi^3$ boson,
and yields the threshold correction to the K\"ahler potential in the $N=2$ sector
(which is identical in form to the gauge threshold correction). Note that there is no contribution to the one-loop K\"ahler potential in the $N=2$ sector for the $\Phi^1,\Phi^2$ fields.

This is the same structure that was found for the study of gauge threshold corrections \cite{09014350}. The Yukawa couplings run
not from the string scale, but instead from the winding scale: the amplitude (\ref{wywy}) diverges logarithmically
until winding states are included in the partition function, which only happens for small values of $t \sim (R M_s)^{-2}$.
More specifically, tapdole cancellation requires that near $t = 0$, $Z(t) \to 0$ with $Z(t) \sim e^{-\frac{1}{R^2 t}}$.
However in the large $t$ regime $Z(t) \sim 1 + \mc{O}(e^{-R^2 t})$. The key cross-over point is $t \sim 1/R^2$,which regulates
field theory running at a scale $E \sim M_W$.

This can be associated to a coupling of the Yukawa to the N=2 twisted sector field that can propagate away from the singularity.
As with the threshold corrections, finiteness does not occur until this tadpole is cancelled.\footnote{Note that in F-theory set-ups of \cite{Donagi:2008ca,Conlon:2009qa} hypercharge flux is typically used for doublet-triplet splitting and so restricts non-trivially to the Higgs matter curves. Since the hypercharge flux is globally trivial it is associated to an $N=2$ sector and so the Higgs fields couple to $N=2$ sectors and therefore the winding scale running should be present for such models.} Finally note that in this case the running was pure $N=2$ and this is because in the absence of orientifolds the $N=1$ running is proportional to the $N=1$ tadpoles (or non-Abelian anomalies) which are cancelled locally. If orientifolds were present there would also be $N=1$ contributions to the running which would terminate at the string scale rather than the winding scale due to their local nature.

The other interesting limit of the amplitude is the case where $t \to \infty$ and $|z_i - z_j|$ remain large. Writing $z_1-z_2=ixt/2$ we have,
for $x$ between 0 and $1/2$,
\begin{align}
\label{phase}
f(x,\theta_1) =& \frac{\vt^\prime (0) \vt (xit/2 + \theta_1)}{\vt(xit/2) \vt(\theta_1)} \nn\\
=& -i \pi\coth \pi xt/2 + \pi\cot \pi \theta_1 + 4\pi \sum_{m,n=1}^\infty e^{-\pi mnt} \sin(\pi m xit +2\pi n \theta_1) \nn \\
\to & -i \pi + \pi \cot \pi \theta \;.
\end{align}
There is no dependence on the vertex operator locations and the non-trivial parts of the integral are
\be
k_2 \cdot k_3 \int \frac{dt}{t^3} Z(t)\int dz_1 dz_2 dz_3 \hbox{Tr}_L(t^1 t^2 t^3 \theta^K) \hbox{Tr}_R(\theta^K) e^{-k_1\cdot k_2 \mc{G}(z_{12})
-k_1\cdot k_3 \mc{G}(z_{13}) - k_2 \cdot k_3 \mc{G}(z_{23})} \;. \label{eq82}
\ee
The crucial physics is contained in this integral. Let us first discuss the overall understanding of the physics before proceeding to the more detailed aspects. Since the correlators $\mc{G}$ go like the distance between the $z_i$ and these are taken of the order of $t$ the integral takes the schematic form
\be
k^2 \int_0^{\infty} dt \; e^{-t k^2} \;.
\ee
This integral evaluates to a finite constant values which is the finite vertex renormalisation we have been discussing. To see this consider the different ranges of the values of $t$. For $t \ll \frac{1}{k^2}$ we have that the exponential is essentially 1 and therefore the contribution to the integral goes like $k^2 t$ which is very small. Near the range $t \sim \frac{1}{k^2}$ the intergal gets an order 1 contribution. In the limit $t \gg \frac{1}{k^2}$ the integrand vanishes due to the exponential factor and so again there is no substantial contribution to the integral. Therefore we see that the renormalisation effect comes from modes around $t \sim \frac{1}{k^2}$. Since in the on-shell limit $k^2 \rightarrow 0$ this is a strict IR effect. This is a sign that this term, while real, should be associated with the 1PI action: it requires the existence of massless particles and comes from the $k \to 0$ limit of the loop integral.

This is exactly what we expect from the field theory. To extract the field theory limit of the amplitude, we use that as $t\rightarrow \infty$
\be
\label{bbb}
G(z_{ji}) \rightarrow - 4 \pi \ap(z_j - z_i) \left( 1 - \frac{2 \hbox{Im}(z_j - z_i)}{t} \right) + ...\;,
\ee
for $\hbox{Im}(z_j - z_i)$ lying between $0$ and $t/2$. The ellipses deonte terms that do not depend on the $z_i$. Since $z_1 > z_2 > z_3$ we use the conformal Killing vector on the torus to set $z_3=0$, then write $z_{1} = (it/2)x_1, z_2 = (it/2) x_2, t =T/2\pi \ap$. We obtain
\be
(k_2 \cdot k_3) \hbox{Tr}_L(t^1 t^2 t^3 \theta^K) \hbox{Tr}_R(\theta^K) \int \frac{dT}{2\pi\ap}  \int_0^1 dx_2 \int_{x_2}^1 dx_1 \exp [ -T (2 k_1 \cdot k_2 (1-x_1) x_2 + k_1^2 x_1(1-x_1) + k_2^2 x_2(1-x_2) )].
\label{jackdaw}
\ee
A change of variable $x_1 \to 1- x_1$ then almost exactly reproduces the
expression (\ref{RawWest}), up to a slightly different momentum prefactor.
To evaluate (\ref{jackdaw}) as an on-shell amplitude, we should impose an infrared energy cutoff
on the $T$ integral of $1/k_2 \cdot k_3$, and then take the limit where $k_2 \cdot k_3 \to 0$.

There are a number of subtleties here. The first concerns
the fact that the Chan-Paton trace is modified as vertex operators are passed through it ($\hbox{Tr}(t^1 \theta^K t^2 t^3) \neq
\hbox{Tr}(t^1 t^2 t^3 \theta^K)$). It is simplest to deal with this by putting the orbifold twist operator at $0$.
In this case the additional phase introduced by the Chan-Paton trace is cancelled by the phase in (\ref{phase}),
which also varies by $e^{2 \pi i \theta}$ as the coordinate $x$ is brought around the annulus.

The second concerns terms in (\ref{bbb}) that are independent of
$z$. For on-shell amplitudes, these do not affect the exponential correlators
$\langle \prod_i e^{ik \cdot X} \rangle$ as their contribution to the exponent
automatically vanishes as $(\sum k_i)^2 = 0$. For closed string off-shell amplitudes, it is argued
in \cite{Minahan} (see p56 journal numbering) that modular invariance of the amplitude
still requires $\sum_{i,j} k_i \cdot k_j = 0$. While modular invariance is not a defining feature of open string amplitudes,
we assume that such terms likewise do not contribute in the open string exponentials.

The third subtlety is concerned with the deep IR limit $t \gg \frac{1}{k^2}$. Although in our schematic argument above we claimed that in this limit the bosonic correlators exponential vanishes this is not strictly the case. The reason is that in for example (\ref{jackdaw}) we can take $t \gg \frac{1}{k^2}$ but also say $x_1 \to 0$ such that the exponent remains finite. This corresponds to bringing the vertices closer together although they are still seperated by scales of order $\frac{1}{k^2}$ which is large. The problem is that this limit is not really appropriate within the 3-point computation because we are probing scales that are only meaningful in an off-shell calculation. The schematic discussion above still holds becasue these are not the scales which are contributing to the constant term in the 1PI action. This can be seen intuitively by noting that it is not possible to have such contributions in the strict $k^2 \to 0$ limit as there is no region $t > 1/k^2$. Indeed in the next section we will be able to probe this contribution by calculating a 4-point amplitude and show that these modes are giving rise to Sudakov logarithms that are associated to scattering amplitudes.

Even taking into account these subtleties there is still
an off-shell ambiguity in this regularisation, which presents itself in the difference of the momentum prefactor; performing the calculation with the PCOs acting on different operators will yield the same amplitude but with a different prefactor. It is simply related to the ambiguity of the off-shell extension for exactly three-point amplitudes (it is unambiguous for two, and for four or more we may go on shell). This and the other ambiguities can all be resolved using a 4-point function, to which we now turn.

\subsection{4-point amplitude}
\label{sec:4ptamp}

\begin{figure}
\begin{center}
\epsfig{file=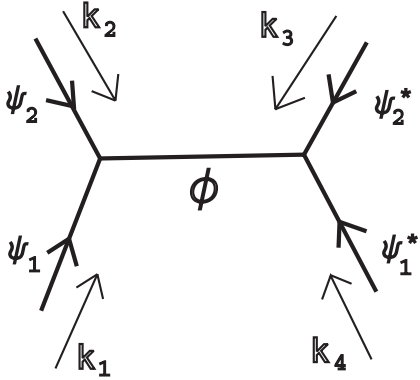, height=5cm}
\caption{The tree-level process which we want to compute the loop corrected amplitude for.}
\label{treeyukawa}\end{center}\end{figure}

To resolve the off-shell ambiguities encountered in the 3-point function we now compute the on-shell 4-point function of
two chiral matter fermions and their partners in an $N=2$ sector, taking the shifts to be $(\theta, -\theta,0)$.
This diagram has unambiguous momentum prefactors and so we can use it to confirm the results of the 3-point calculation while resolving
the off-shell ambiguities.
We are interested in the loop level computation of the process whose tree-level structure is given by figure \ref{treeyukawa}.
The Chan-Paton factors are
\be
t_{\psi_1} = \left( \begin{array}{cccc} 0 & 1 & 0 & 0 \\ 0 & 0 & 0 & 0 \\ 0 & 0 & 0 & 0 \\ 0 & 0 & 0 & 0 \end{array} \right), \quad
t_{\psi_2} = \left( \begin{array}{cccc} 0 & 0 & 0 & 0 \\ 0 & 0 & 1 & 0 \\ 0 & 0 & 0 & 0 \\ 0 & 0 & 0 & 0 \end{array} \right), \quad
t_{\psi_2^{*}} = \left( \begin{array}{cccc} 0 & 0 & 0 & 0 \\ 0 & 0 & 0 & 0 \\ 0 & 1 & 0 & 0 \\ 0 & 0 & 0 & 0 \end{array} \right), \quad
t_{\psi_1^{*}} = \left( \begin{array}{cccc} 0 & 0 & 0 & 0 \\ 1 & 0 & 0 & 0 \\ 0 & 0 & 0 & 0 \\ 0 & 0 & 0 & 0 \end{array} \right).
\ee
This requires the vertex operators to be ordered as $\psi_1(z_1) \psi_2(z_2) \psi_2^{*}(z_3) \psi_1^{*}(z_4)$. As we wish to factorise
the 4-point diagram onto a 3-point Yukawa interactions we will be interested in the limit that $z_1 \to z_2$ (there is another limit
$z_3 \to z_4$ which, for the spin structure taken below, factorises the diagram onto the exchange of a gauge boson). We also require a single trace amplitude with all vertex
operators on the same end of the cylinder.

As the 4-point amplitude is evaluated at finite momentum and with all particles on-shell we can without loss of generality require that
$\psi_1$ has momentum $(k,-k,0,0)$, implying that $k_1^{1+} = k_1^{2+} = k_1^{2-} = 0$.
The Dirac equation $\Gamma \cdot k |\psi \rangle = 0$ then
fixes the H-charges of the $-1/2$-picture vertex operator to be $(+,+,+,-,-)$. The H-charges of the
operators are given by
\bea
\psi_1 & : & (+,+,+,-,-), \nonumber \\
\psi_2 & : & \alpha_2 (-,-,-,+,-) + \beta_2 (+,+,-,+,-) , \nonumber \\
\psi_2^{*} & : & \alpha_3 (+,-,+,-,+) + \beta_3 (-,+,+,-,+), \nonumber \\
\psi_1^{*} & : & \alpha_4 (-,+,-,+,+) + \beta_4 (+,-,-,+,+).
\eea
The coefficients $\alpha_i, \beta_i$ are determined by the momenta $k_2$, $k_3$ and $k_4$.. However in the limit we are interested in
$(k_1 \cdot k_2 \ll 1)$, then $\alpha_2 =  1 + \mc{O}(k_1 \cdot k_2)$ and $(\alpha_3, \beta_3) \cdot (\alpha_4, \beta_4) =
1 + \mc{O}(k_1 \cdot k_2)$. This limit corresponds to both the $(\psi_1, \psi_2)$ and $(\psi_3, \psi_4)$ pairs
having momenta that are essentially back to back.
In this case it is
straightforward to evaluate the amplitude, choosing to picture change the two chiral fermions $\psi_1$ and $\psi_2$.

We first consider the case of picture changing the internal direction, e.g.
\be
\begin{array}{c} \psi_1(z_1) \\ \psi_2(z_2) \\ \psi_2^{*}(z_3) \\ \psi_1^{*}(z_4) \end{array} \qquad
\left( \begin{array}{ccccc} + & + & + & - & - \\ - & - & - & + & - \\ + & - & + & - & + \\ - & + & - & + & + \end{array} \right)
\longrightarrow \left( \begin{array}{ccccc} + & + & - & - & - \\ - & - & + & + & - \\ + & - & + & - & + \\ - & + & - & + & + \end{array} \right)
\ee
It is straightforward to show that all these type of contributions vanish after Riemann summation and so we should only consider external picture
changing.

The external picture changing also turns out to be straightforward. In fact there are only two non-trivial cases, coming from
\be
\begin{array}{c} \psi_1(z_1) \\ \psi_2(z_2) \\ \psi_2^{*}(z_3) \\ \psi_1^{*}(z_4) \end{array} \qquad
\left( \begin{array}{ccccc} + & + & + & - & - \\ - & - & - & + & - \\ + & - & + & - & + \\ - & + & - & + & + \end{array} \right)
\longrightarrow \left( \begin{array}{ccccc} +++ & + & + & - & - \\ --- & - & - & + & - \\ + & - & + & - & + \\ - & + & - & + & + \end{array} \right)
\label{term1}
\ee
and
\be
\begin{array}{c} \psi_1(z_1) \\ \psi_2(z_2) \\ \psi_2^{*}(z_3) \\ \psi_1^{*}(z_4) \end{array} \qquad
\left( \begin{array}{ccccc} + & + & + & - & - \\ - & - & - & + & - \\ - & + & + & - & + \\ + & - & - & + & + \end{array} \right)
\longrightarrow \left( \begin{array}{ccccc} +++ & + & + & - & - \\ --- & - & - & + & - \\ - & + & + & - & + \\ + & - & - & + & + \end{array} \right)
\label{term2}
\ee
There is no analogous combination with the PCOs acting on the second external direction because $k_1^{2\pm}=0$. All other possibilities end up vanishing. Typically, without the inclusion of derivative terms the terms vanish due to the Riemann summation with a double zero.
In principle derivative terms can then lead to a non-zero answer. However, the fact that there is a double zero implies that
we need derivative terms for both $\psi_1(z_1)$ and $\psi_2(z_2)$.
However derivative terms for $\psi_1(z_1)$ always involve a contraction of (say) $\partial X^{1+}(z_1) e^{ik_1 \cdot X(z_1)}$, which brings down a factor of $k_1^{1+}$ and therefore vanishes by the equations of motion.

For (\ref{term1}) the fermionic spin sums end up giving
\be
4k_1 \cdot k_2 \frac{\vartheta_1(z_1 - z_2 + z_3 - z_4 + \theta) \vartheta_1(-\theta) \vartheta_1(z_1 - z_3) \vartheta_1(z_2 - z_4) \eta^6}
{\vartheta_1(z_1 - z_2) \vartheta_1(z_1 - z_4) \vartheta_1(z_2 - z_3) \vartheta_1(z_3 - z_4) \vartheta_1(\theta) \vartheta_1(-\theta)},
\label{bang1}
\ee
and for (\ref{term2}) the fermionic spin sums give
\be
4k_1 \cdot k_2 \frac{\vartheta_1(z_1 - z_2 + \theta) \vartheta_1(-z_3 + z_4 - \theta) \eta^6}{\vartheta_1(z_1 - z_2) \vartheta_1(z_3 - z_4) \vartheta_1(\theta)
\vartheta_1(-\theta)}.
\label{bang2}
\ee
We have used the fact that $k_1^{1-}$ is the only non-vanishing part of $k_1$ to write $k_1^{1-} k_2^{1+}$ as $2k_1 \cdot k_2$\footnote{Actually there is an additional Lorentz structure present that we are neglecting here for clarity, since it cannot contribute a correction to a Yukawa vertex. We discuss the complete structure in appendix \ref{APPENDIX:LORENTZ}.}.
(\ref{bang1}) and (\ref{bang2}) have the same structure as $z_1 \to z_2$ or $z_3 \to z_4$, which are the two limits which factorise the diagram
onto 3-point Yukawa interactions. This reflects the fact that the Yukawa involves the exchange of a spin-0 scalar boson which carries no
helicity structure.
However (\ref{bang1}) also has poles as $z_1 \to z_4$ and $z_2 \to z_3$, which are absent in (\ref{bang2}). These poles correspond to
the exchange of a spin-1 gauge boson exchange. This is helicity-forbidden from appearing in (\ref{bang2}).

In any case, when we take $z_1 \to z_2$ to factorise onto a Yukawa diagram, we obtain the expression
\be
\label{hermes}
\int \frac{dt}{t^3} Z(t) \int dz_1 dz_3 dz_4 \frac{k_1 \cdot k_2}{k_1 \cdot k_2} \frac{\vartheta_1(z_3 - z_4 + \theta) \vartheta_1^{'}(0)}
{\vartheta_1(z_3 - z_4) \vartheta_1(\theta)} \langle \prod_i e^{i k \cdot X} \rangle.
\ee
There are two interesting limits of this expression. The first is to take $z_3 \to z_4$ (this is limit C of figure \ref{BigDiagram}).
This generates a pole $1/k_3 \cdot k_4$ and gives an amplitude of the form
\be
\frac{1}{k_1 \cdot k_2} \int \frac{dt}{t^3} Z(t) \int dz_1 dz_3 \langle \prod_i e^{ik_i \cdot X} \rangle \to
\frac{1}{k_1 \cdot k_2} \int_{t=1/m_W^2}^{t \sim 1/k_1 \cdot k_2} \frac{dt}{t}.
\ee
This gives the logarithmic correction to the scalar propagator, with the ultraviolet cut off at the winding scale from the partition
function, and the infrared regulated by $\langle e^{ik \cdot X} \rangle$ at an energy scale $1/t \sim k_1 \cdot k_2$.
This matches the running we found in section \ref{sec:3ptamp} for the 3-point amplitude which we refer to for further discussion regarding this physics. The pole in $(k_1\cdot k_2)$ is associated to the scalar propagator in the 4-point diagram, and so should be abstracted before evaluating the
Yukawa coupling.

The other interesting limit is to take $t \to \infty, |z_i - z_j| \sim \mc{O}(t)$. In this limit we have,
\be
\int \frac{dt}{t^3} \int dz_1 dz_3 dz_4 \hbox{Tr}(t^1 t^2 t^3 t^4 \theta^K) (-i \pi + \pi \cot \pi \theta)
\exp \left[ -s G(z_{34}) - t (G(z_{13}) + G(z_{24})) - u (G(z_{14}) + G(z_{23}) ) \right]. \label{eq94}
\ee
where $s= k_1 \cdot k_2$, $t = k_1 \cdot k_3$, $u= k_1 \cdot k_4$, $z_{ij} = z_i - z_j$ and
\be
\label{luke}
G(z_{ji}) = - 4 \pi \hbox{Im}(z_j - z_i) \left( 1 - \frac{2 \hbox{Im}(z_j - z_i)}{t} \right) + ... \;,
\ee
for $\hbox{Im}(z_j - z_i)$ lying between $0$ and $t/2$. The ellipses in (\ref{luke}) denote terms that do not depend on the $z_i$.
 When inserted in (\ref{eq94}), with the $z_{12}$ correlator included, such terms drop out due to the $(s+t+u)=0$ factor multiplying them.
As $z_1 \simeq z_2$ up to very small corrections, we can use $s + t + u = 0$ to simplify (\ref{eq94}) to
\be
\label{sky}
\int \frac{dt}{t^3} \int dz_1 dz_3 dz_4 \hbox{Tr}(t^1 t^2 t^3 t^4 \theta^K) (-i \pi + \pi \cot \pi \theta)
\exp \left[ -s G(z_{34}) + s( G(z_{13}) + G(z_{14}) ) \right].
\ee
We can further simplify this in two ways. First, we do a coordinate redefinition $z \to \frac{2z}{it}$ so that the $z$ coordinates
now run from $0$ to 1 rather than 0 to $it/2$. Secondly, we can use
(\ref{luke}) to simplify the exponent of (\ref{sky}). Doing so we find that we can write
\be
\label{sky2}
\frac{(-i \pi + \pi \cot \pi \theta)}{8 i} \int dt \int_{0}^{1} dz_1 dz_3 dz_4 \, \hbox{Tr}(t^1 t^2 t^3 t^4 \theta^K)
\exp \left[ -4s \pi t z_{31}(1- z_{41}) \right].
\ee
Written in this form $z_{31}$ and $z_{41}$ must take values between 0 and 1.

The integral (\ref{sky2}) contains all the relevant physics. Let us begin by noting how this recreates the physics found in the 3-point calculation of section \ref{sec:3ptamp} and how it resolves any amiguities associated to going off-shell. First, as discussed above, there is an extra pole in $s$ coming from the scalar propagator in the scattering which should be extracted before comparing with the 3-point Yukawa coupling calculation. This effectively means that we should multiplty (\ref{sky2}) by $s$ and compare it to (\ref{eq82}). We see that they take the same schematic form. However now the momentum factor in the exponential $s$ and the prefactor, also $s$, match and cancel exactly solving the momentum factors mismatch of the 3-point calculation. For scales $t \lesssim \frac{1}{s}$ all the physics is the same and we find again the finite renormalisation of the 1PI superpotential. This serves as a non-trivial check of our calculations and physics understanding.


However in a scattering amplitude the loop integral also involves the deep IR region $t \gg \frac{1}{s}$. Such contributions are a necessary
 component of vertex renormalisation, as the physical amplitude requires a sum over virtual and real components. Viewed as part of a 4-point scattering amplitude, (\ref{sky2}) is (as one would expect) infrared divergent with Sudakov logarithms.

 We now see this
explicitly by evaluating (\ref{sky2}).
In principle the vertex operators in (\ref{sky2}) can take
any ordering, but the Chan-Paton traces enforce an ordering $(z_1, z_3, z_4)$ and cyclic permutations of this.
To evaluate (\ref{sky2}) we then take the three orderings,
$0 < z_1 < z_3 <z_4 < 1$, $0 < z_4 < z_1 < z_3 < 1$ and $0< z_3 < z_4 < z_1 < 1$, and evaluate each in turn, focusing on the
behaviour for large values of $t$.

The first case $0 < z_1 < z_3 <z_4 < 1$ gives
\bea
& & \int dt \int_{0}^1 dz_4 \int_0^{z_4} dz_1 \int_{z_1}^{z_4} dz_3 \exp \left[ -4s \pi t (z_3 - z_1) (1- (z_4 - z_1)) \right] \nonumber \\
& = & \int dt \int_0^1 dz_4 \int_0^{z_4} dz_1 \frac{1 - \exp \left[ - 4 \pi s t(z_4 - z_1)(1- (z_4 - z_1)) \right]}{4 \pi s t(1 - (z_4 - z_1))} \nonumber \\
& \simeq & \int dt \int_0^1 dz_4 \frac{1}{4 \pi st } \left[ \ln (1 + z_1 - z_4) \right]_0^{z_4 - \mc{O}(1/st)}.
\eea
The modified upper limit is due to the effective cutoff that occurs for $z_1 = z_4 - \mc{O}(1/st)$, as
$(1 - \exp \left[-4 \pi s t(z_4 - z_1)(1-z_4 + z_1) \right]) \to 0$.
\bea
\int dt \int_0^1 dz_4 \frac{1}{4 \pi st } \left[ \ln (1 + z_1 - z_4) \right]_0^{z_4 - \mc{O}(1/st)} & = & \int dt \int_0^1 dz_4 \frac{-\ln (1 - z_4)}{4 \pi s t} \nonumber \\
& = & \int \frac{dt}{4 \pi s t} \to \frac{1}{4 \pi s} \ln \left( \frac{s}{\mu^2} \right),
\eea
for $\mu^2 < s$ acting as an infrared energy cutoff.

The second case $0 < z_4 < z_1 < z_3 < 1$ gives
\bea
& & \int dt \int_0^1 dz_3 \int_0^{z_3} dz_1 \int_0^{z_1} dz_4 \exp \left[ -4 \pi s t (z_3 - z_1)(z_1 - z_4) \right] \nonumber \\
& = & \int dt \int_0^1 dz_3 \int_0^{z_3} dz_1 \frac{1 - \exp \left[ - 4 \pi s t(z_3 - z_1) z_1 \right]}{4 \pi s t(z_3 - z_1)}.
\eea
The exponential term is only relevant for $z_1 \lesssim 1/st$ and $z_3 - z_1 \lesssim 1/st$, where it causes the integrand to vanish.
So in effect we can write
\bea
\int dt \int_0^1 dz_3 \int_0^{z_3} dz_1 \frac{1 - \exp \left[ - 4 \pi s t(z_3 - z_1) z_1 \right]}{4 \pi s t(z_3 - z_1)}
& \simeq & \int dt \int_{1/st}^{1} dz_3 \int_{1/st}^{z_3 - 1/st} dz_1 \frac{1}{4 \pi s t(z_3 -z_1)} \nonumber \\
& = & \int dt \int_{1/st}^1 dz_3 \frac{1}{4 \pi s t} \left[ \ln (st) + \ln z_3 \right] \\
& = & \int dt  \frac{\ln (st) - 1}{4 \pi st} \to \frac{1}{4 \pi s} \left( \frac{1}{2} \ln^2 \left( \frac{s}{\mu^2} \right) - \ln \left( \frac{s}{\mu^2} \right) \right), \nonumber
\eea
with an infrared cutoff of $\mu^2$. This has the characteristic form of a Sudakov double logarithim.

The final case gives
\bea
\mc{A_C} & = & \int dt \int_0^1 dz_1 \int_0^{z_1} dz_3 \int_{z_3}^{z_1} dz_4 \exp \left[ - 4 \pi s t (1 + (z_3 - z_1))(z_1 - z_4) \right] \nonumber \\
& = & \int dt \int_0^{1} dz_1 \int_0^{z_1} dz_3 \frac{1 - \exp \left[-4 \pi st (1 + z_3 - z_1)(z_1 - z_3) \right]}{4 \pi s t(1 + z_3 - z_1)} \nonumber \\
& \simeq & \int dt \int_0^1 dz_1 \frac{-\ln(1-z_1)}{4 \pi s t} \nonumber \\
& = & \int dt \frac{1}{4 \pi s t} \to \frac{1}{4 \pi s} \ln \left( \frac{s}{\mu^2} \right).
\eea

We should now sum all three terms with each having the same sign. As in the 3-point amplitude and the discussion
below eq. (\ref{jackdaw}), there is a phase factor from the
Chan-Paton trace that is cancelled by a phase from the term $\frac{\vartheta_1(z_3 - z_4 + \theta)}{\vartheta_1(z_3
-z_4) \vartheta_1(\theta)}$ in (\ref{hermes}).

We see that the dominant contribution comes from the second case, and has the structure of a Sudakov double logarithim. This is a
(field theory) infrared divergence which is a necessary part of a vertex renormalisation
in a scattering amplitude. The divergence is physical, and
can be regulated by including diagrams involving a tree-level vertex but with additional soft real emission below
the scale of the infrared cutoff. The sum of the virtual and real soft contributions then cancels the infrared divergence,
leaving a finite result for the probability of scattering without extra particle emission.
Note that this expression is not identical to (\ref{RawWest}), as there is not a single effective Feynman vertex
that can be inserted into a scattering diagram - consistency instead requires the presence of the above infrared
divergence in the vertex that has to be regulated by the inclusion of unobservable real emission below a given
resolution scale.

In summary then, the 4-point amplitude confirms and extends the physics of the 3-point amplitude, while removing the
off-shell ambiguities. With on-shell momenta the 3-point amplitude can only have trivial kinematics, and thus can give
purely a correction to the `1PI superpotential'. In the 4-point amplitude, we can place all particles on-shell
while keeping a finite momentum transfer. In this case we obtain vertex renormalisation with non-trivial kinematics
(as the momenta entering the vertex are off-shell). As necessary for a consistent field theory interpretation, this
vertex renormalisation is infrared divergent, with the divergence being regulated by the inclusion of unobservable
soft real emission.

\section{Summary and discussion}
\label{sec:summary}

In this paper we studied the behaviour of one-loop Yukawa couplings in a local type IIB model
by calcluating both 3-point and 4-point amplitudes on the annulus. Both amplitudes recreated the same physics.
In particular, we showed that the Yukawas are renormalised by both wavefunction renormalisation and vertex renormalisation.

Wavefunction renormalisation can be incorporated into the Wilsonian action as running of the kinetic terms.
Threshold corrections for the kinetic terms were shown
to give Yukawa running up to the winding scale rather than the local string scale.
This is the same as occurred for gauge coupling running and the underlying
reasons are the same: the Yukawas couple to a closed string $N=2$ mode that locally
sources a tadpole but which must be cancelled for the runnning to stop. This can only happen once the closed string mode
reaches the bulk radius: alternatively, that winding modes are incorporated into the open string computation.
This effect is important both conceptually, as for a truly non-compact local model the Yukawas would recieve infinite threshold corrections,
and phenomenologically. For example it shows that in local GUT models
bottom-tau mass unification occurs at the same (winding) scale as that at which gauge unification occurs (even though this scale
is parametrically above the string scale).

We also showed that vertex renormalisation occurs at one-loop.
This is an infrared - but physical - field theory effect, and is absent
from previous studies of one-loop Yukawas in string theory.\footnote{We expect that it should be present also within IIA and Heterotic setting.}
It arises in the infrared limit of the annulus from loops containing light degrees of freedom.
It matches work done for supersymmetric field theories which argued that such an effect can be present in theories with massless particles \cite{West,JackJonesWest,West:1991qt}.
The field theory interpretation of this effect is as an $\int \frac{D^2}{\square} g(\Phi)$
operator.  The renormalisation does not affect the Wilsonian superpotential as it is associated to a loop integral over
light modes but can be understood as a 1PI `effective superpotential'.

It would be interesting to exploit this effect for phenomenological purposes.
One natural potential application of this is to supersymmetric models of radiative flavour generation.
Models of Yukawa couplings in string theory often suffer from a rank-one problem, with corrections to this structure being
only non-perturbatively small. For the models studied here, there is a correction to pre-existing Yukawa couplings. However
it would be very interesting to determine whether or not new Yukawa couplings could be generated in this way at one-loop.
If this was the case then an immediate application would be to study flavour physics with radiative Yukawa generation in the supersymmetric phase. Even if the renormalisation is purely that of existing Yukawa couplings it should still be important to take it into account.

\subsection*{Acknowledgments}

We thank Marcus Berg, Michael Dine and Emilian Dudas for helpful discussions.

The work of EP was supported in part by the European ERC Advanced Grant 226371 MassTeV, by the CNRS PICS no. 3059 and 4172,
by the grants ANR-05-BLAN-0079-02, the PITN contract PITN-GA-2009-237920 and the IFCPAR CEFIPRA programme 4104-2. MDG is supported by the German Science Foundation (DFB) under SFB 676. JC is supported by
the Royal Society with a University Research Fellowship and Balliol College, Oxford.

\appendix

\section{Gauge Threshold Corrections}
\label{sec:gaugethres}
We here describe the computation of gauge threshold corrections for vector bosons,
previously carried out in the background field
formalism \cite{09014350, 09061920}. We redo this using vertex operators and compute the one-loop `scattering'
of two gauge bosons, from which we can extract the propagator. This is as in e.g. \cite{Marcus1, Marcus2, Abel:2003ue,Abel:2006hk}.
Other studies of gauge threshold corrections for IIA/IIB brane models in string theory include \cite{9906039, 0302221,
0612234, 07052150, 07110866, 09100843}.
As string theory is an on-shell theory this formally vanishes due to kinematic prefactors ($k^2 = 0$) but the threshold
corrections can be extracted by removing this kinematic term.

The computation is carried out on the annulus and requires the insertion of two gauge boson vertex operators. These are
given in the (-1) and 0-pictures by
\bea
\mc{V}^{-1} & = & t^a \psi^{\mu} e^{i k \cdot X} \\
\mc{V}^{0} & = & t^a \left( i \frac{d X^{\mu}}{d \tau} + 2 \alpha' \left( k \cdot \psi \right) \psi^{\mu} \right) e^{i k \cdot X}
\eea
Here $t^a$ is the Chan-Paton factor. The annulus requires an overall ghost charge of zero and so we need both vertex operators
in the 0 picture.  We furthermore
need to sum over all possible annular spin structures. The amplitude we want is
\be
\mc{A}_{vv} = g_s^2 \int_0^{\infty} \frac{dt}{(8 \pi^2 \alpha' t)^2 t} \int  dz_1 dz_2 \sum_{\alpha} c_{\alpha}
\langle V_0(z_1 ;k_1, a_1) V_0 (z_2; k_2, a_2) \rangle.
\ee
We have included the factors $(8 \pi^2 \alpha' t)^{-2}$ coming from the partition function over the external momentum modes.
Here $\alpha$ represents the spin structure, $z_i$ the insertion point, $k_i$ the momentum and $a_i$ the polarisation. The amplitude is
understood as being evaluated in the orbifold and so there is also a trace over Chan-Paton factors.
We take the generators to be non-Abelian and so the structure of a $\theta^k$ orbifold sector is
\be
\mc{A}_{vv}^k = \hbox{Tr}_L (t^a t^b \theta^k) \hbox{Tr}_R (\theta^k) \mc{A}
\ee
where $\hbox{Tr}_{L,R}$ denotes a trace over the left and right-handed ends of the string.
The basic CFT correlator required is
$$
\sum_{\alpha} \langle \left( \partial X^{\mu} - 2 i \alpha' (k_1 \cdot \psi) \psi^{\mu} \right) e^{i k_1 \cdot X} \left( \partial X^{\nu}
-2 i \alpha'  (k_2 \cdot X) \psi^{\nu} \right) e^{i k_2 \cdot X} \rangle
$$
Cross-contractions will vanish as $\langle :(k \cdot \psi) \psi^{\nu}: \rangle = 0$ due to normal ordering, and so we are left with two basic correlators
to evaluate
\be
\label{first}
\langle \partial X^{\mu} e^{i k_1 \cdot X (z_1)} \partial X^{\nu} e^{i k_2 \cdot X(z_2)} \rangle
\ee
and
\be
\label{second}
\langle (k_1 \cdot \psi) \psi^{\mu} e^{i k_1 \cdot X(z_1)} (k_2 \cdot \psi) \psi^{\nu} e^{i k_2 \cdot X(z_2) }
\ee
We can evaluate these by contracting terms using either the bosonic or fermionic Green's functions until we have
simply the identity operator. The expectation value of the identity operator gives the partition function, which depends both on
the spin structure of the annulus and also on the modular parameter.

The first case of eq. (\ref{first}) involves only bosonic correlators. Bosonic fields are insensitive to the
spin structure, and so
when contracting all the bosons  we will be left with an expression of the form
$$
\mc{B}(z,w) \sum_{\alpha} \langle 1 \rangle_{\alpha}.
$$
This is a sum over the partition functions for each spin structure, and therefore vanishes in a supersymmetric background.

To obtain a non-vanishing amplitude we instead consider the case (\ref{second}) which involves contractions of the fermionic fields.
We can contract fermions using eq. (\ref{hchargecorr}), which gives for the even spin structures in the external directions
$$
S_{\alpha}(z-w) = \frac{\vartheta_{\alpha}(z-w) \vartheta_1^{'}(0)}{\vartheta_1 (z-w) \vartheta_{\alpha}(0)}.
$$
Note the extra factor of $\vartheta_{\alpha}(0)$ in the deonominator compared to (\ref{hchargecorr}): the reason is that here
we have not yet taken into account the fermionic partition function (which gives an extra factor of $\vartheta_{\alpha}(0)$ in the numerator)
wherea in (\ref{hchargecorr}) this term is already included.

The odd spin structure (P,P) is not
relevant for this amplitude as we can never saturate the fermionic zero modes.
We can now evaluate the amplitude (\ref{second}). We can (effectively)
split (\ref{second}) up into
bosonic and fermionic correlators,
\be
\langle e^{ik \cdot X(z_1)} e^{i k \cdot X(z_2)} \rangle \langle (k \cdot \psi) \psi^{\mu} (z_1)
(k \cdot \psi) \psi(z_2) \rangle_{\alpha}.
\ee
The bosonic correlator gives
$$
\langle e^{ik_1 \cdot X(z_1)} e^{ik_2 \cdot X(z_2)} \rangle = \exp \left[- k_1 \cdot k_2 \mc{G}(z_1 - z_2) \right].
$$
Let us define $f_1^{\mu \nu} = k_1^{\mu} a_1^{\nu} - k_1^{\nu} a_1^{\mu}$ and similarly for $f_2$. The fermionic correlator
then gives
\be
\label{fermcol}
\langle (k \cdot \psi) \psi^{\mu} (z_1)
(k \cdot \psi) \psi(z_2) \rangle_{\alpha} = (f_1 f_2) S_{\alpha}^2(z_1 - z_2) \langle 1 \rangle_{\alpha}
\ee
Note that for an on-shell gauge boson $k_1 \cdot k_2 = 0$. However this is really a kinematic prefactor associated to the
presence of an $F_{\mu \nu} F^{\mu \nu}$ term, and so we can formally extract it and obtaing the threshold corrections from
the remainder of the amplitude.

The fermionic terms need to be summed over spin structures. We can simplify the amplitude using the following
results (see e.g. \cite{Bianchi:2006nf}).
\be
S_{\alpha}^2(z-w) = P(z-w) - e_{\alpha - 1},
\ee
where
\be
P(z) = - \partial_z^2 \ln \vartheta_1(z) - \frac{1}{3} \frac{ \vartheta_1^{'''}(0)}{\vartheta_1^{'}(0)}, \qquad
e_{\alpha - 1} = -4 \pi i \partial_{\tau} \ln \left( \frac{\vartheta_{\alpha}(0, \tau)}{\eta(\tau)} \right).
\ee
As $P(z)$ is independent of $\alpha$, it follows that after summing over spin structures the only non-zero term
comes from $e_{\alpha - 1}$, and specifically from the $\partial_{\tau} \ln \vartheta_{\alpha}(0, \tau)$ part of
$e_{\alpha - 1}$ (as all other terms give a pure sum over partition functions, which vanishes). So the only relevant part of
$S_{\alpha}^2(z-w)$ is
$$
e_{\alpha - 1} = - 4 \pi i \left( \frac{\partial_{\tau} \vartheta_{\alpha}(0, \tau)}{\vartheta_{\alpha}(0, \tau)} \right)
+ (\hbox{terms independent of $\alpha$}).
$$
Note that this is also independent of $(z-w)$, and so the amplitude summed over spin structures is independent of
the insertion points. This is fortunate as it dramatically simplifies the integrals.

We now evaluate (\ref{fermcol}). The partition function in an $\mc{N}=1$ twisted sector is
\be
\langle 1 \rangle_{\mc{N}=1} = \frac{ \vartheta_{\alpha}(0)}{\eta^3} \prod_i \frac{\vartheta_{\alpha}(u_I)}{\vartheta_1(u_I)}
\ee
So combining the fermionic propagator and the partition function we have
$$
\sum_{\alpha} \eta_{\alpha} \frac{\partial_{\tau} \vartheta_{\alpha}(0, \tau)}{\vartheta_{\alpha}(0, \tau)}
\frac{\vartheta_{\alpha}(0, \tau)}{\eta^3} \prod_i \frac{\vartheta_{\alpha}(u_i)}{\vartheta_1(u_i)}
$$
where $\eta_{\alpha}$ is the spin structure phase factor.
Now as $\partial_{\tau} \vartheta_{\alpha} = \partial_z^2 \vartheta_{\alpha}$, for $\mc{N}=1$ and $\mc{N}=2$ sectors
this simplifies to
\bea
\mc{N}=1 & : & \sum_{\alpha} \frac{\vartheta_{\alpha}^{''}(0)}{\eta^3} \prod_i \frac{\vartheta_{\alpha}(u_i)}{\vartheta_1(u_i)}, \\
\mc{N}=2 & : &
\sum_{\alpha} \frac{\vartheta_{\alpha}^{''}(0)}{\eta^3} \frac{\vartheta_{\alpha}(0)}{\eta^3}
\frac{\vartheta_{\alpha}(u_i)}{\vartheta_1(u_i)} \frac{\vartheta_{\alpha}(-u_i)}{\vartheta_1(-u_i)}.
\eea
This is now precisely the same expressions that occurred when we were studying the threshold corrections via the
backgound field method.

The Chan-Paton factors are straightforward to evaluate. In each sector, we obtain a factor of
$
\hbox{Tr}_L (T^a T^b \theta^K) \hbox{Tr}_R(\theta^K).
$
In the $\theta$ and $\theta^3$ sectors
the traces vanish once anomaly cancellation is imposed, whereas the $\theta^2$ sector gives the form of the beta functions.
\bea
\hbox{Tr}_L (T^a T^b \theta) \hbox{Tr}_R(\theta) & \sim & (n_0 - n_2), \\
\hbox{Tr}_L (T^a T^b \theta^2) \hbox{Tr}_R(\theta^2) & \sim & (n_0 - n_1).
\eea

The amplitude
\be
\sum_{\theta^K} \int \frac{dt}{t^3} \int dz_1 dz_2 \exp \left[- k_1 \cdot k_2 \mc{G}(z_1 - z_2) \right] (f_1 f_2) S_{\alpha}^2(z_1 - z_2) \langle 1 \rangle_{\alpha}
\ee
can then be restricted to the $\mc{N}=2$ sector. Furthermore, in the limit $k^2 \to 0$ we neglect the term $ \exp \left[- k_1 \cdot k_2 \mc{G}(z_1 - z_2) \right]$ and extract the kinematic prefactor $f_1 f_2 \to F_{\mu \nu} F^{\mu \nu}$.
The overall amplitude is then
\be
\mc{A} = (f_1 f_2) \int_0^{\infty} \frac{dt}{t^3} \int dz_1 dz_2 (n_0 - n_2)
\sum_{\alpha} \frac{\vartheta_{\alpha}^{''}(0)}{\eta^3} \frac{\vartheta_{\alpha}(u_i)}{\vartheta_1(u_i)} \frac{\vartheta_{\alpha}(-u_i)}{\vartheta_1(-u_i)} Z(t).
\ee
Here $Z(t)$ is the partition function for winding strings charged under the gauge group.
As there is now no $z$ dependence in the integral, the integration over the $z_i$ coordinates simply gives two powers of $t$,
giving an amplitude
\be
\mc{A}_K = (f_1 f_2) \int_0^{\infty} \frac{dt}{t} \left( \hbox{Tr}_L \left( T^a T^b \theta^K \right) \hbox{Tr}_R (\theta^K)
\sum_{\alpha} \frac{\vartheta_{\alpha}^{''}(0)}{\eta^3} \prod_i \frac{\vartheta_{\alpha}(u_I)}{\vartheta_1(u_I)} \right) Z(t).
\label{finalgauge}
\ee
Up to the formally vanishing prefactor $f_1 f_2$, this is precisely the expression for threshold corrections previously
obtained using the background field formalism \cite{09014350}.

\section{Theta Identities}

The standard notation for the Jacobi Theta functions is:
\begin{equation}
\vartheta \left[ \begin{array}{c} a \\ b \end{array} \right] (z;\tau) = \sum_{n = -\infty}^{\infty} \mathrm{exp} \bigg[ \pi i (n+a)^2 \tau + 2 \pi i (n+a)(z+b) \bigg]
\end{equation}
A common definition is $\vartheta_{\alpha \beta} \equiv \vartheta \left[ \begin{array}{c} \alpha/2 \\ \beta/2 \end{array} \right]$, and
\begin{eqnarray}
\vartheta_1 \equiv \vartheta_{11} & \vartheta_2 \equiv \vartheta_{10} \nonumber \\
\vartheta_3 \equiv \vartheta_{00} & \vartheta_4 \equiv \vartheta_{01}.
\end{eqnarray}

Expansions of the functions for $q=e^{\pi i \tau}$ are
\begin{eqnarray}
\vartheta_{00} (z, \tau) & = \vartheta_3 =& 1 + 2\sum_{n=1}^{\infty} q^{n^2} \cos 2\pi n z \nonumber \\
\vartheta_{01} (z, \tau) & = \vartheta_4 =& 1 + 2\sum_{n=1}^{\infty} (-1)^n q^{n^2} \cos 2\pi n z \nonumber \\
\vartheta_{10} (z, \tau) & = \vartheta_2 =& 2q^{1/4}\sum_{n=0}^{\infty} q^{n(n+1)} \cos \pi (2n+1) z \nonumber \\
\vartheta_{11} (z, \tau) & = \pm \vartheta_1 =& 2q^{1/4}\sum_{n=0}^{\infty} (-1)^n q^{n(n+1)} \sin \pi (2n+1) z \nonumber \\
\end{eqnarray}

The Dedekind $\eta$ function is defined as
\begin{eqnarray}
\eta (\tau) &=& q^{1/12} \prod_{m=1}^{\infty} (1-q^{2m}) \\
&=& \left[ \frac{\vartheta_1^{\prime} (0,\tau)}{-2\pi} \right]^{1/3}.
\end{eqnarray}

The generalised Riemann summation formula is
\begin{align}
&\sum_{\alpha, \beta} (-1)^{\alpha + \beta + \alpha \beta}\prod_{i=1}^{4} \tab{\alpha/2+c_i}{\beta/2+d_i}{z_i}{\tau}=2\tab{1/2}{1/2}{\sum_i z_i/2}{\tau} \tab{1/2 +c_2 }{1/2 + d_2 }{\frac{z_1 + z_2 - z_3-z_4}{2} }{\tau}\nonumber \\
&\times \tab{1/2 +c_3 }{1/2 + d_3 }{\frac{z_1 - z_2 + z_3-z_4}{2} }{\tau}\tab{1/2 +c_4 }{1/2 + d_4 }{\frac{z_1 - z_2 - z_3+z_4}{2} }{\tau} \label{rieiden1}.
\end{align}

We also have the five-theta identity \cite{Atick:1986rs}:
\begin{align}
&\sum_\nu \delta_\nu \vartheta_\nu (z_1) \vartheta_\nu (z_2) \vartheta_\nu (z_3) \vartheta_\nu (z_4) \vartheta_\nu (z_5) \vartheta_\nu^{-1} (z_1 + z_2 + z_3 + z_4 + z_5) \nonumber \\
&=-2 \vartheta_1 (z_1 + z_2 + z_3 + z_4) \vartheta_1 (z_2 + z_3 + z_4 + z_5) \vartheta_1(z_1 + z_3 + z_4 + z_5) \nonumber \\
&\times \vartheta_1 (z_1 + z_2 + z_4 + z_5)\vartheta_1 (z_1 + z_2 + z_3 + z_5) \vartheta_1^{-1} (2[z_1 + z_2 + z_3 + z_4 + z_5]).
\label{FiveTheta}\end{align}

Another identity useful for threshold corrections is
\be
\sum_{\alpha} \frac{\vartheta_{\alpha}^{''}(0)}{\eta^3} \prod_i \frac{\vartheta_{\alpha}(u_I)}{\vartheta_1(u_I)} =
- 2 \pi \sum_{i=1}^3 \frac{\vartheta^\prime \left[ \begin{array}{c} 1/2 \\ 1/2 \end{array}\right](\vartheta_i,it)}{\tab{1/2}{1/2}{\vartheta_i}{it}}.
\label{extrathetaidentity}
\ee

\section{Determination of Amplitude via Lorentz Structure }
\label{APPENDIX:LORENTZ}

We are seeking to compute the four-fermion amplitude
\begin{align}
\int_0^\infty \frac{dt}{(4\pi^2 \ap t)^2} \frac{2}{it} \bigg(\int_0^{it/2} \prod_{i=1}^4 dz_i\bigg) \mathcal{A}_4 \equiv& \\
 \int_0^\infty \frac{dt}{(4\pi^2 \ap t)^2} &\frac{2}{it} \bigg(\int_0^{it/2} \prod_{i=1}^4 dz_i\bigg) N u^\alpha_1 u^\beta_2 v^{\dot{\gamma}}_3 v^{\dot{\delta}}_4 \lim_{u\rightarrow z_1} \lim_{v\rightarrow z_2} \bra \dot{X}_\mu (u) \dot{X}_\nu (v) \prod_{i=1}^4 e^{ik_i \cdot X(z_i)} \ket \nn\\
&\times (u-z_1)^{1/2} (v-z_2)^{1/2} \bra S_\alpha (z_1) S_\beta (z_2) \tilde{S}_{\dot{\gamma}} (z_3) \tilde{S}_{\dot{\delta}} (z_4) \psi^\mu (v)\psi^\nu (u) \ket\nn
\end{align}
where we are picture-changing the two chiral operators; the $u_\alpha^i, v_{\dot{\gamma}}^i$ are the wavefunctions for the chiral and anti-chiral fermions respectively. $N$ is a normalisation factor absorbing the normalisation of the PCOs and vertex operators determined at tree level. In general it is necessary to also have amplitudes where the PCOs act on the internal directions, but it can be easily shown that in this case all such amplitudes vanish.

The fermionic portion of the amplitude must have the structure
\begin{align}
\bra S_\alpha (z_1) S_\beta (z_2) \tilde{S}_{\dot{\gamma}} (z_3) \tilde{S}_{\dot{\delta}} (z_4) \psi^\mu (v)\psi^\nu (u) \ket =& A \hat{C}_{\alpha \beta} \hat{C}_{\dot{\gamma}\dot{\delta}} \eta^{\mu\nu} + B (\hat{C}\Gamma^\mu\Gamma^\nu)_{\beta \alpha} \hat{C}_{\dot{\gamma}\dot{\delta}} \nonumber \\
&+C (\hat{C}\Gamma^\nu\Gamma^\mu)_{\dot{\delta}\dot{\gamma}} \hat{C}_{\alpha \beta} + D(\hat{C}\Gamma^\nu \Gamma^\rho)_{\dot{\delta}\dot{\gamma}} (\hat{C}\Gamma^\mu \Gamma_\rho)_{\beta \alpha}.
\end{align}
This is the same as in \cite{Atick:1987gy}, where a similar computation was performed in the context of the Heterotic string. We can determine $A, B, C, D$ as functions of the vertex operator positions by computing a judicious choice of string amplitudes. For this, we work in a helicity basis, with $\{\Gamma^{a+}, \Gamma^{b-}\} = \delta^{ab}, \{\Gamma^{a+}, \Gamma^{b+}\} = \{\Gamma^{a-}, \Gamma^{b-}\} = 0$, spinors $| a - 1/2, b - 1/2 \rangle = (\Gamma^{2+})^b (\Gamma^{1+})^a | - - \rangle$ corresponding to the spin field $e^{i(a-1/2)H_1 } e^{i(b-1/2)H_2}$ and have the charge conjugation matrix $\hat{C}$ where
\begin{align}
\hat{C}|--\ket =& |++\ket\nn\\
\hat{C}|++\ket =& -|--\ket\nn\\
\hat{C}|+-\ket =& -|-+\ket\nn\\
\hat{C}|-+\ket =& |+-\ket.
\end{align}

Then calculating the amplitude $A_1$ where $\alpha = |++\ket, \beta = |--\ket, \dot{\gamma} = |+-\ket,\dot{\delta} = |-+\ket$, we have for $\mu=1+,\nu=1-$
\beq
A_1 \equiv \frac{1}{2}A -B.
\eeq
If we then have $\mu=1-,\nu=1+ $ we find
\beq
A_2 \equiv \frac{1}{2} A -C.
\eeq
Further choosing $ \mu=2+,\nu=2- $ gives
\beq
A_3 \equiv \frac{1}{2}A -B -C + 2D
\eeq
and $\mu=2-,\nu=2+ $
\beq
A_4 \equiv \frac{1}{2} A.
\eeq
This allows easy determination of the coefficients:
\begin{align}
A =& 2A_4\nn\\
B=& A_4 - A_1 \nn\\
C=& A_4 - A_2 \nn\\
D=& \frac{1}{2} (A_3 - A_2 - A_1 +A_4).
\end{align}

Computation of these four amplitudes then yields the whole result; any other choice of spinors or $\mu,\nu$ must be consistent with this. Once we compute these four amplitudes for an $N=2$ sector of the theory, we find that once we take spin-structure summation and the equations of motion into account only two are non-vanishing. This is because the enhanced symmetry of this sector causes there to be no bosonic derivative terms contributing; all of the amplitudes are thus contracted with $k_1, k_2$ and we find terms $B$ and $D$ vanish by the equations of motion. Thus we need only consider two amplitudes, $A_2$ and $A_4$.

To determine $A_2$ we compute the fermionic correlator with charges
\begin{align}
\psi_1^{+1/2}(z_1) =& \half(+++,+,+,-,-) \;, \nn \\
\psi_2^{+1/2}(z_2) =& \half(---,-,-,+,-) \;, \nn \\
\tilde{\psi}_3^{-1/2}(z_3) =& \half(+,-,-,+,+) \;, \nn \\
\tilde{\psi}_4^{-1/2}(z_4) =& \half(-,+,+,-,+) \;.
\end{align}
Afer Riemann summation we obtain
\begin{align}
X_2 \equiv 2\frac{\vt(z_1 - z_2 + \theta)\vt(z_3 - z_4 - \theta)(\vartheta_1^\prime(0))^2}{\vt(z_1 -z_2) \vt(z_3 - z_4)\vt(\theta)\vt(-\theta)}
\end{align}

To determine $A_4$ we compute the fermionic correlator  with charges
\begin{align}
\psi_1^{+1/2}(z_1) =& \half(+,+++,+,-,-) \;, \nn \\
\psi_2^{+1/2}(z_2) =& \half(-,---,-,+,-) \;, \nn \\
\tilde{\psi}_3^{-1/2}(z_3) =& \half(+,-,-,+,+) \;, \nn \\
\tilde{\psi}_4^{-1/2}(z_4) =& \half(-,+,+,-,+) \;.
\end{align}
Afer Riemann summation we obtain
\begin{align}
X_4 \equiv 2\frac{\vt(z_1 - z_2 - z_3 + z_4+ \theta)\vt(z_1 - z_4) \vt(z_2 - z_3)(\vartheta_1^\prime(0))^2}{\vt(z_1 -z_2) \vt(z_3 - z_4)\vt(z_1 - z_3) \vt(z_2 - z_4)\vt(\theta)}.
\end{align}

So then the total result is
\begin{align}
\mathcal{A}_4 =& (2\ap)^2 N\bigg[ 2X_4 k_1 \cdot k_2 (u_1 \cdot u_2) (u_3 \cdot u_4) + (X_4 - X_2) (\tilde{u}_4 C\slashed{k}_1 \slashed{k}_2 \tilde{u}_3) (u_1 \cdot u_2)\bigg] Z(t) \nn\\
=& (2\ap)^2 N\bigg[ (X_2 + X_4) k_1 \cdot k_2 (u_1 \cdot u_2) (u_3 \cdot u_4) + (X_4 - X_2) k_{1\mu} k_{2\nu} (\tilde{u}_4 \hat{C}\Gamma^{\mu\nu} \tilde{u}_3) (u_1 \cdot u_2)\bigg] Z(t).
\end{align}
where $u_1 \cdot u_2 = u^\alpha_1 C_{\alpha\beta} u_2^\beta$, $\Gamma^{\mu\nu } \equiv \frac{1}{2} [\Gamma^\mu,\Gamma^\nu]$ and $Z(t)$ is the bosonic classical partition function in the untwisted torus (\ref{CompactUntwistedPF}).

\end{document}